\documentclass[12pt]{article}
\usepackage{epsf,epsfig,amssymb,amsmath,graphicx,float,latexsym} 

\newcommand{\lsim}{\raisebox{-0.13cm}{~\shortstack{$<$ \\[-0.07cm] $\sim$}}~}
\newcommand{\gsim}{\raisebox{-0.13cm}{~\shortstack{$>$ \\[-0.07cm] $\sim$}}~}

\begin{document}
\renewcommand{\thefootnote}{\fnsymbol{footnote}}

\begin{titlepage}
  
\begin{flushright}
June 2010
\end{flushright}

\begin{center}

\vspace{1cm}

{\Large {\bf Direct and Indirect Detection of Neutralino Dark\\
\vspace*{1mm}
 Matter and Collider Signatures in an $SO(10)$ Model \\
\vspace*{2mm}
 with Two Intermediate Scales}} 

\vspace{1cm}

{\sc Manuel Drees}$^{a\,}$\footnote{drees@th.physik.uni-bonn.de},
{\sc Ju Min Kim}$^{a\,}$\footnote{juminkim@th.physik.uni-bonn.de}, and
{\sc Eun-Kyung Park}$^{a,b\,}$\footnote{park@lapp.in2p3.fr}\\

\vskip 0.15in
{\it 
$^a${Physikalisches Institut and Bethe Center for Theoretical Physics,
  Universit\"at Bonn, Nussallee 12, 53115 Bonn, Germany}\\
$^b${LAPTH, Universit\'e de Savoie, CNRS, \\
B.P. 110, F-74941 Annecy-le-Vieux Cedex, France}
}
\vskip 0.5in

\abstract{We investigate the detectability of neutralino Dark Matter
  via direct and indirect searches as well as collider signatures of
  an $SO(10)$ model with two intermediate scales. We compare the
  direct Dark Matter detection cross section and the muon flux due to
  neutralino annihilation in the Sun that we obtain in this model with
  mSUGRA predictions and with the sensitivity of current and future
  experiments. In both cases, we find that the detectability improves
  as the model deviates more from mSUGRA. In order to study collider
  signatures, we choose two benchmark points that represent the main
  phenomenological features of the model: a lower value of $|\mu|$
  and reduced third generation sfermion masses due to extra Yukawa
  coupling contributions in the Renormalization Group Equations, and
  increased first and second generation slepton masses due to new
  gaugino loop contributions. We show that measurements at the LHC can
  distinguish this model from mSUGRA in both cases, by counting events
  containing leptonically decaying $Z^0$ bosons, heavy neutral Higgs
  bosons, or like--sign lepton pairs. }

\end{center}
\end{titlepage}
\setcounter{footnote}{0}

\section{Introduction}

Grand Unified Theories (GUTs) based on the gauge group $SO(10)$ have
been considered good candidates for the unification of electroweak and
strong interactions \cite{oldso10}. All matter fields of one
generation are incorporated in a single irreducible representation,
the spinor {\bf 16}. Moreover, the ``seesaw'' mechanism \cite{seesaw},
which can explain small neutrino masses as indicated by neutrino
oscillations, is naturally embedded.

In \cite{Drees:2008tc}, two of us in particular chose a model by
Aulakh et al. \cite{Aulakh:2000sn}, a supersymmetric $SO(10)$ model
with two intermediate scales: $SO(10)$ is first broken to $SU(4)_C
\times SU(2)_L \times SU(2)_R$ by a {\bf 54} dimensional Higgs at the
GUT scale $M_X$; then to $SU(3)_C \times U(1)_{B-L} \times SU(2)_L
\times SU(2)_R$ by {\bf 45} at scale $M_C$; finally to the Standard
Model gauge group by ${\bf 126}+{\bf \overline{126}}$ at scale
$M_R$. Imposing the unification condition for the gauge couplings
fixes the intermediate scales $M_C$ and $M_R$ for given $M_X$;
i.e. $M_X$ is a free parameter. However, its lower bound is set by the
lower bound on the lifetime of the proton \cite{pdg}. We took $M_X =
3\cdot 10^{15}$GeV as default value. In addition, a second pair of
Higgs doublets was introduced, in order to modify the minimal $SO(10)$
predictions for the masses of quarks and leptons, which are not
consistent with experiments. In order to compare the low energy
phenomenology of the model with that of mSUGRA \cite{msugra}, we
assumed universal soft breaking parameters ($m_0, M_{1/2}, A_0$) as
boundary condition at the GUT scale.

The {\bf 126}--dimensional Higgs whose vacuum expectation value breaks
$SU(2)_R \times U(1)_{B-L}$ to $U(1)_Y$ also gives Majorana masses to
the right--handed neutrinos. The resulting masses for the light
neutrinos are schematically written as
\begin{eqnarray} \label{seesaw}
m_\nu = \frac{m_D^2}{M_N} = \frac{(Y_{\nu}\langle
  H_u^0\rangle)^2}{Y_N\langle {\bf \overline{126}}\rangle}. 
\end{eqnarray}
The Yukawa coupling $Y_N$, and to a lesser extent $Y_\nu$, gives new
contributions to the Renormalization Group Equations (RGEs) of the
MSSM Yukawa couplings and soft breaking parameters. Therefore the
weak--scale masses, and thus the radiative electroweak symmetry
breaking and the relic density of Dark Matter, depend on $Y_N$, and
hence on the light neutrino masses for fixed $M_R$. This remains
qualitatively true for other $SO(10)$ GUTs with a ``type--I'' seesaw
mechanism at an intermediate scale. Note that $Y_\nu$ unifies with the
up--type quark Yukawa couplings, and is hence fixed. For given $M_R$,
and hence given $\langle {\bf \overline{126}}\rangle$, the absolute
scale of the light neutrino masses is thus determined by $Y_N$, with
{\em larger} $Y_N$ yielding {\em lighter} neutrinos. For our minimal
choice $M_X = 3 \cdot 10^{15}$ GeV, the requirement that $Y_N$ remains
perturbative at least up to scale $M_X$ therefore leads to the lower
bound $m_\nu \gsim 0.15$ eV on the mass of the heaviest light
neutrino.

For our current study the most important modification of the
weak--scale spectrum is the reduction of the higgsino mass parameter
$|\mu|$, which comes about as follows. The weak--scale stop masses are
reduced compared to the mSUGRA prediction, due to the Yukawa coupling
given by \cite{Drees:2008tc}
\begin{eqnarray}
W_{\rm Yuk, 422} \ni \frac{1}{2} Y_N \left(F^c \overline{\Sigma}_R F^c
  + F \overline{\Sigma}_L F \right) \,. 
\end{eqnarray}
Here $W_{\rm Yuk,422}$ is the superpotential valid between the scales
$M_C$ and $M_X$, $\overline{\Sigma}_R$ and $\overline{\Sigma}_L$ are
in the {(${\bf 10}, {\bf 1, 3}$)} and {($\overline{\bf 10}, {\bf 3,
    1}$)} representation, respectively, of the gauge group $SU(4)_C
\times SU(2)_L \times SU(2)_R$, and $F$ and $F^c$ denote quark and lepton
superfields in the {\bf(4,2,1)} and {($\overline{\bf 4}, {\bf 1, 2}$)}
representation. This reduces the term $\propto Y_t^2$ in the RGE for
$m_{H_u}^2$, leading to an increase of $m_{H_u}^2$ at the weak
scale. As a result, electroweak symmetry breaking requires {\em smaller}
values $|\mu|$ than in mSUGRA.

Another distinctive feature of the model is the rapid increase of the
gauge couplings at high energies. This is due to the introduction of
large additional Higgs representations, needed in order to break the
gauge symmetry. As a result, relations between weak--scale and the
GUT--scale soft breaking parameters are modified
\cite{Drees:2008tc}. In particular, for a given universal gaugino mass
$M_{1/2}$ at the GUT scale, the $SO(10)$ model predicts much smaller
gaugino masses at the weak scale than mSUGRA does.

In ref.\cite{Drees:2008tc} it was shown that thermal neutralino Dark
Matter remains viable, although the allowed region of parameter space
is even more highly constrained than in mSUGRA. In the next Section,
we will focus on the detectability of this Dark Matter candidate by
direct and indirect searches. In Sec.~3, we compare signatures at the
LHC between mSUGRA and this $SO(10)$ model for two benchmark
points. Finally, we conclude in Sec.~4.

\section{Direct and indirect detection of Dark Matter}

All results presented in this Section are obtained using a modified
version \cite{Drees:2008tc} of \texttt{SOFTSUSY 2.0}
\cite{Allanach:2001kg} to evaluate the mass spectra at the weak
scale. These are then fed into \texttt{micrOMEGAs 2.2}
\cite{micromegas} to calculate the LSP relic density. If this is found
acceptable, we feed the same low--energy spectrum into
\texttt{DarkSUSY 5.0.2} \cite{Gondolo:2004sc} for the calculation of
various Dark Matter detection rates \cite{jungman}. 

Specifically, we compute: elastic LSP--proton scattering cross
sections due to spin--in\-de\-pen\-dent as well as spin--dependent
interactions; the muon flux resulting from LSP annihilation in the
Sun; and the antiproton flux from LSP annihilation in the halo of our
galaxy. In all cases we compare with the sensitivities of the best
current and/or near--future experiments.  In most cases, we only
consider parameter sets leading to a relic density within two standard
deviations of the value found by combining WMAP data with other
cosmological observations \cite{Komatsu2009}:
\begin{eqnarray}\label{eq:wmap}
  \Omega_{CDM}h^2 = 0.1131 \pm 0.0068 \ \ (2 \sigma \ {\rm range}).
\end{eqnarray}

\begin{figure}[ht]
\begin{center}
\includegraphics[width=8.2cm,height=8.2cm,angle=270]{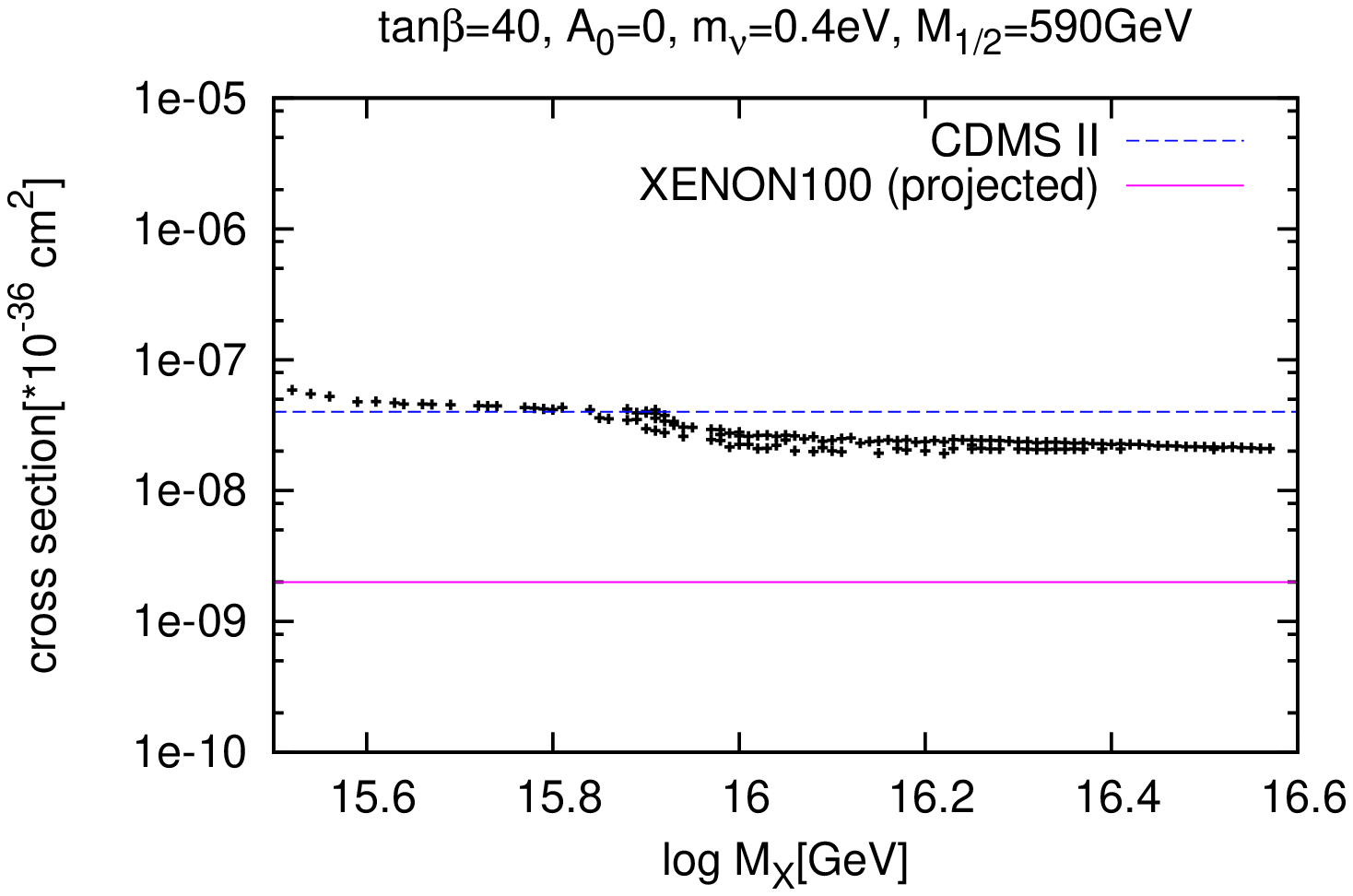}
\includegraphics[width=8.2cm,height=8.2cm,angle=270]{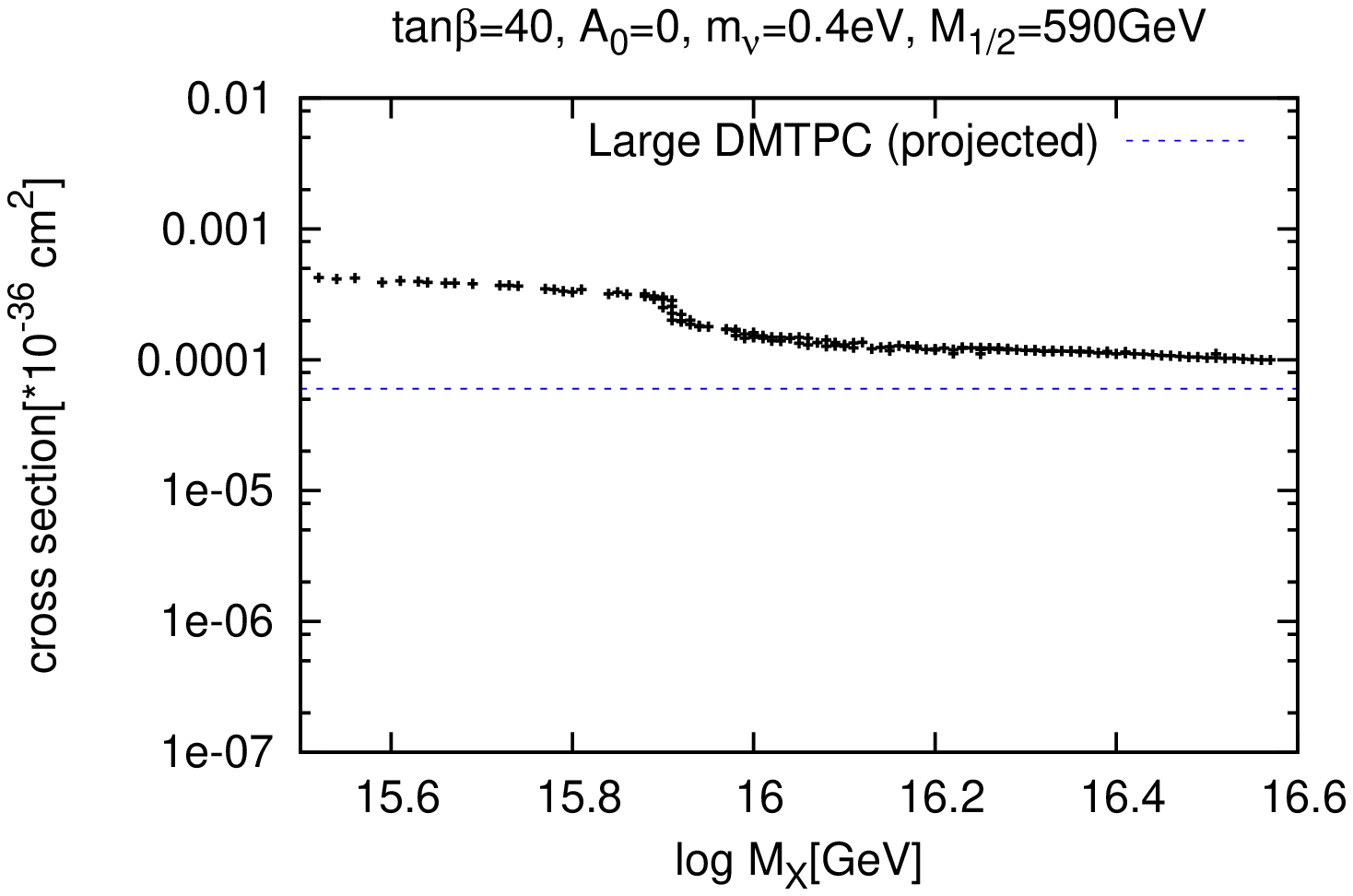}
\caption{The $M_X$ dependence of the SI (left frame) and SD (right
  frame) neutralino--proton scattering cross section. $m_0$ is varied
  such that the relic density satisfies the constraint (\ref{eq:wmap}). The
  neutralino mass $m_{\chi}=124$ GeV, for $M_X=10^{15.5}$ GeV,
  increases to $235$ GeV for $M_X=10^{16.6}$ GeV. The experimental
  constraints are taken for $m_{\tilde\chi_1^0}=120$ GeV.}
    \label{fig:sigmapMx}
\end{center}
\end{figure}

We begin our discussion by analyzing the impact of reducing $M_X$,
i.e. ``switching on'' the intermediate scales, on the elastic
LSP--proton scattering cross section. Figure \ref{fig:sigmapMx} shows
the spin--independent (SI; left frame) and spin--dependent (SD; right
frame) contributions to this cross sections as we vary $M_X$ while
most soft breaking parameters as well as the light neutrino masses are
kept fixed. The scalar mass $m_0$ is varied along with $M_X$ such that
the relic density lies in the range of Eq.(\ref{eq:wmap}). Note that
$m_0^2 \gg M^2_{1/2}$ in this plot, i.e. we are in the region of
significant higgsino--neutralino mixing, which is most favorable for
direct neutralino Dark Matter searches. As a result, the
spin--independent cross section is always well above the projected
sensitivity of the XENON100 experiment \cite{dmtools}, while the
spin--dependent cross section lies above the projected sensitivity of
the DMTPC experiment \cite{dmtools}.

While these gross features remain unchanged, we see that, as $M_X$
decreases, i.e. as the deviation from mSUGRA becomes larger, the cross
section is enhanced, so that for the lowest $M_X$, the
spin--independent cross section slightly exceeds the limit set by the
CDMS II experiment \cite{Ahmed:2009zw}. This is mostly due to the
reduction of the LSP mass for fixed $M_{1/2}$ in the model with
intermediate scales, which we mentioned near the end of Sec.~1. In
particular, for $M_X \leq 8 \cdot 10^{15}$ GeV, $m_{\tilde \chi_1^0} <
m_t$, so that $\tilde \chi_1^0 \tilde \chi_1^0 \rightarrow t \bar t$
annihilation is forbidden. The loss of this important annihilation
channel has to be compensated by increasing bino--higgsino mixing,
i.e. by decreasing $\mu$, which in turn is accomplished by increasing
$m_0$. This leads to increased couplings of the lightest neutralino to
neutral Higgs bosons as well as to the $Z^0$ boson. Note that in
scenarios with gaugino mass unification, first generation squarks are
always much heavier than the lightest neutralino, suppressing their
contributions to LSP--nucleon scattering. Choosing $m_0^2 \gg
M_{1/2}^2$, as done here, further strengthens this hierarchy, so that
Higgs and $Z^0$ exchange contributions largely determine the SI and SD
cross sections, respectively.

The curves in Figs.~\ref{fig:sigmapMx} show a noticeable negative
slope even away from this threshold. In case of the SI cross section,
this is due to the reduction of the mass of the heavier neutral
CP--even Higgs boson with decreasing $M_X$, which goes along with the
reduction of the weak--scale gaugino masses (although for fixed $m_0$
the ratio $m_A / m_{\tilde \chi_1^0}$ slightly increases with
decreasing $M_X$ \cite{Drees:2008tc}). Note also that decreasing
$m_{\tilde \chi_1^0}$ requires a simultaneous, if slower, decrease of
$\mu$, since otherwise the higgsino--component of $\tilde \chi_1^0$
would become too small, yielding too small an annihilation cross
section. This decrease of both weak--scale gaugino masses and of $\mu$
with decreasing $M_X$ implies that the higgsino components of the LSP
become more different in magnitude; note that they become identical in
size for $|\mu| \gg M_Z$. This in turn enhances the $\tilde \chi_1^0
\tilde \chi_1^0 Z^0$ coupling, which is proportional to the difference
of the squares of these components \cite{msugra}.

\begin{figure}[ht]
\begin{center}
\includegraphics[width=8.2cm,height=8.2cm,angle=270]{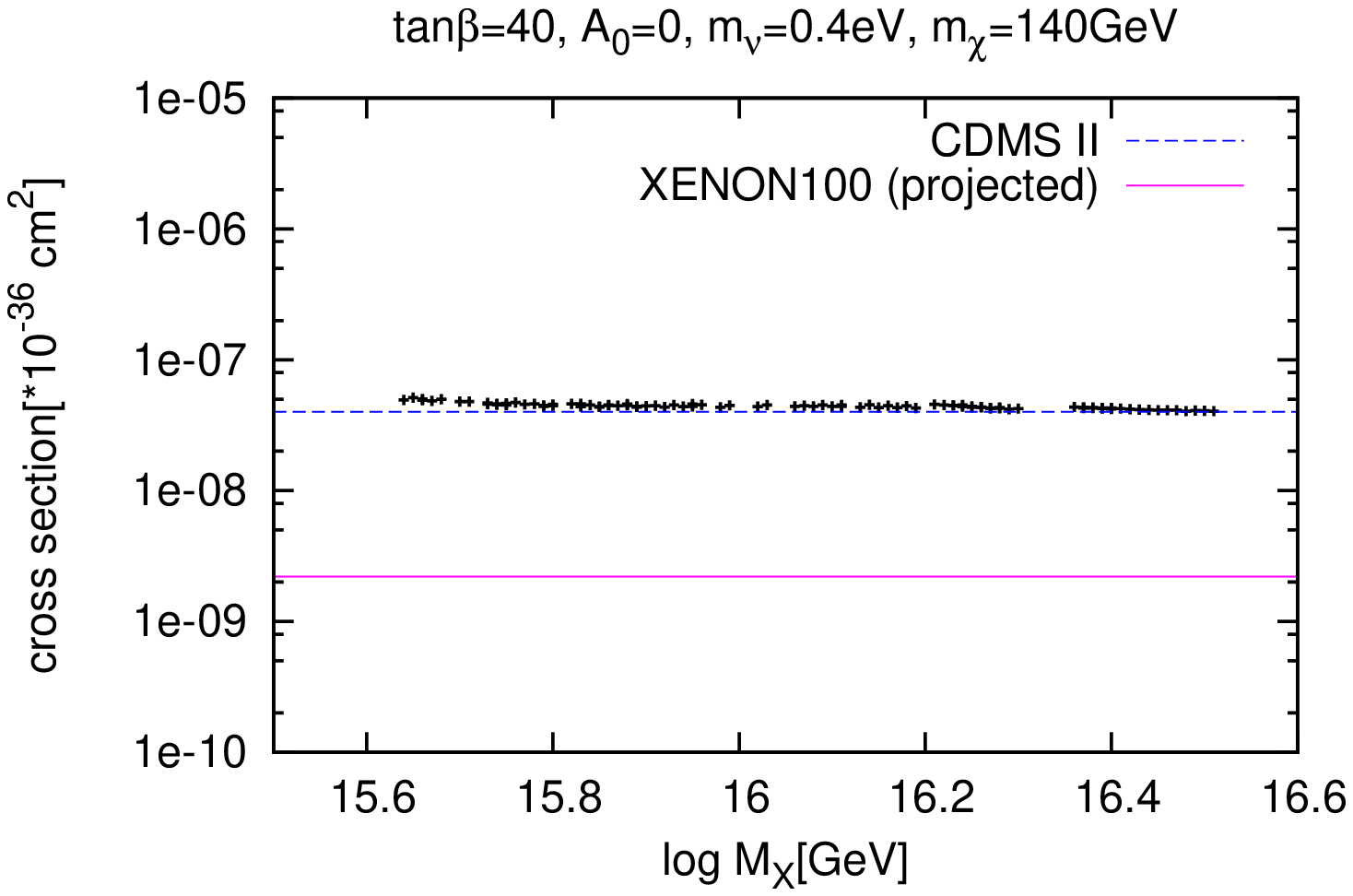}
\includegraphics[width=8.2cm,height=8.2cm,angle=270]{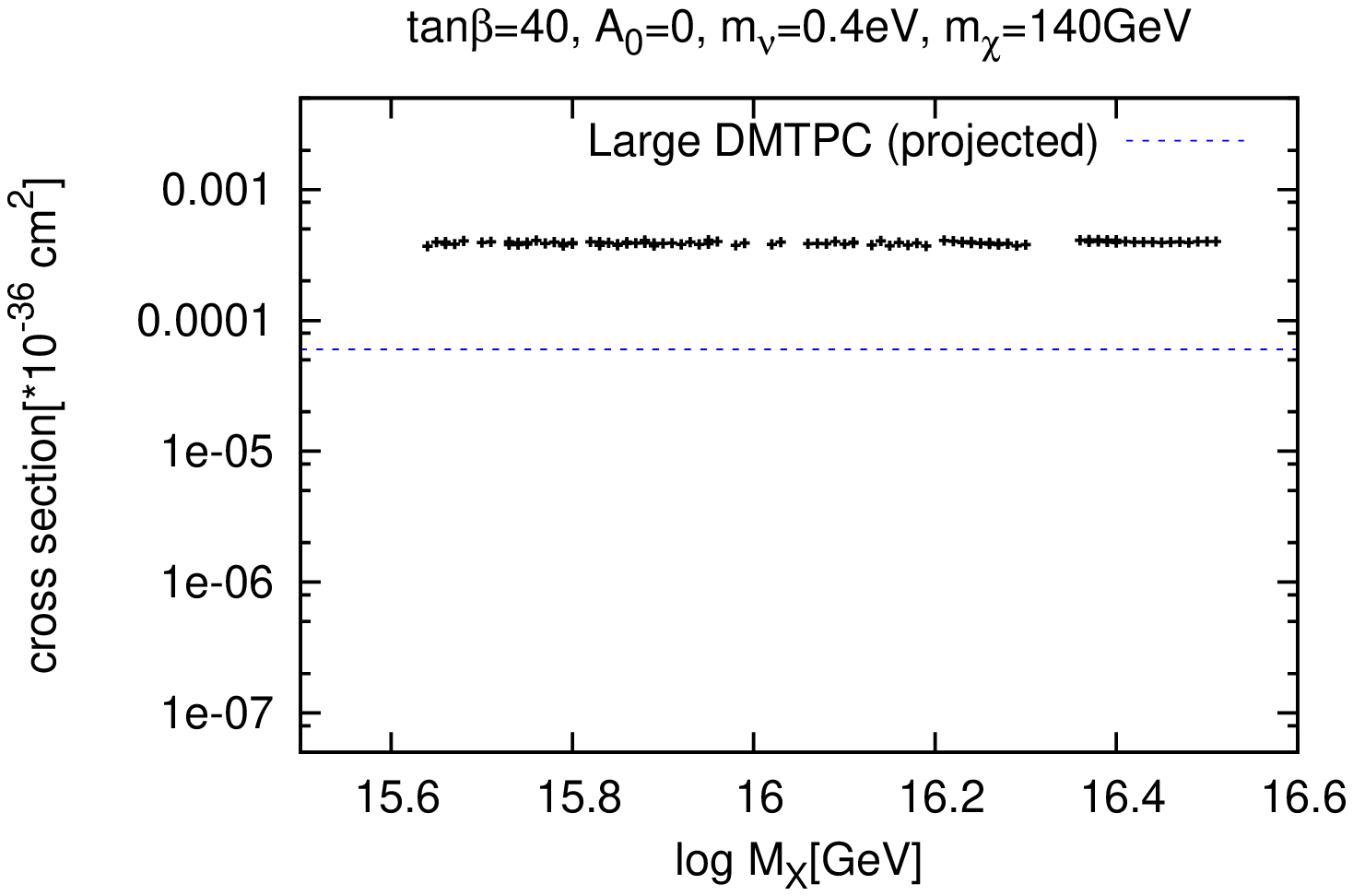}
\caption{As Fig.~\ref{fig:sigmapMx}, except that $M_{1/2}$ has also
  been varied along with $M_X$, such that the mass of the lightest
  neutralino is kept fixed at 140 GeV.}
\label{fig:sigmapMx1}
\end{center}
\end{figure}

Figs.~\ref{fig:sigmapMx1} show results analogous to those in
Fig.~\ref{fig:sigmapMx}, except that now the gaugino mass parameter
$M_{1/2}$ has been varied along with $M_X$ such that the LSP mass is
kept fixed. This required taking larger values of $M_{1/2}$ for
smaller $M_X$. As expected from our previous discussion, the effect of
reducing $M_X$ is now quite small. The spin--independent cross section
(left frame) increases by $\sim 20$\% as $M_X$ is reduced to its
minimal value. This can be explained as follows. Since now $m_{\tilde
  \chi_1^0}$ is kept fixed, we also have to keep $\mu$ essentially
fixed in order to maintain the correct relic density. This requires
reducing $m_0$ when $M_X$ is reduced. This in turn leads to a
reduction of $m_A$, which over--compensates the increase of $m_A$ that
would result if $M_X$ were reduced for fixed $m_0$ and fixed LSP mass.
This implies a similar reduction for the mass of the heavier CP--even
neutral Higgs boson whose exchange plays a prominent role in this
cross section. However, it is not clear whether this variation is
significant given astrophysical and likely experimental
uncertainties. The variation of the spin--dependent cross section is
even smaller.

\begin{figure}[ht]
\begin{center}
\includegraphics[width=8.2cm,height=8.2cm,angle=270]{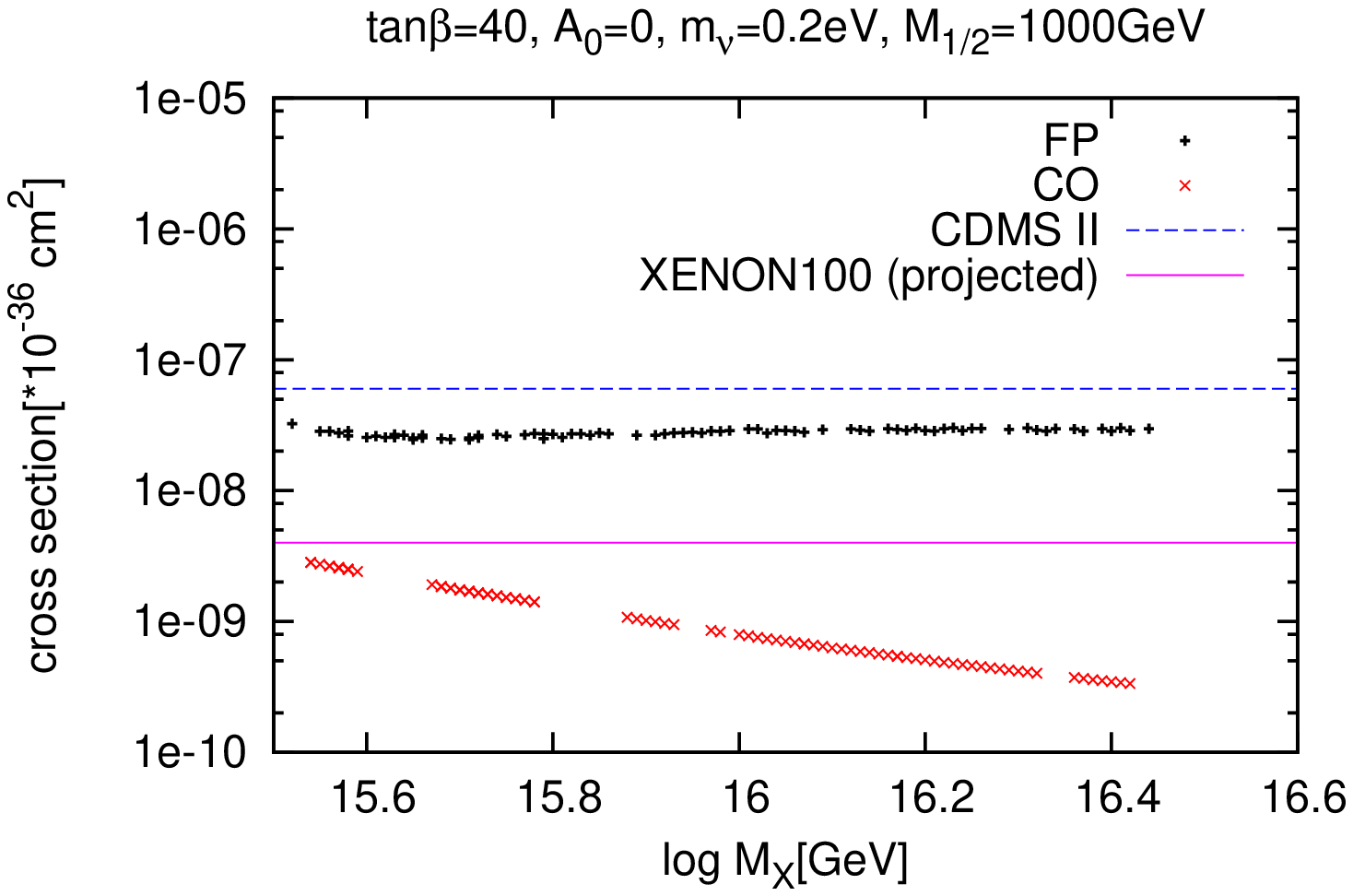}
\includegraphics[width=8.2cm,height=8.2cm,angle=270]{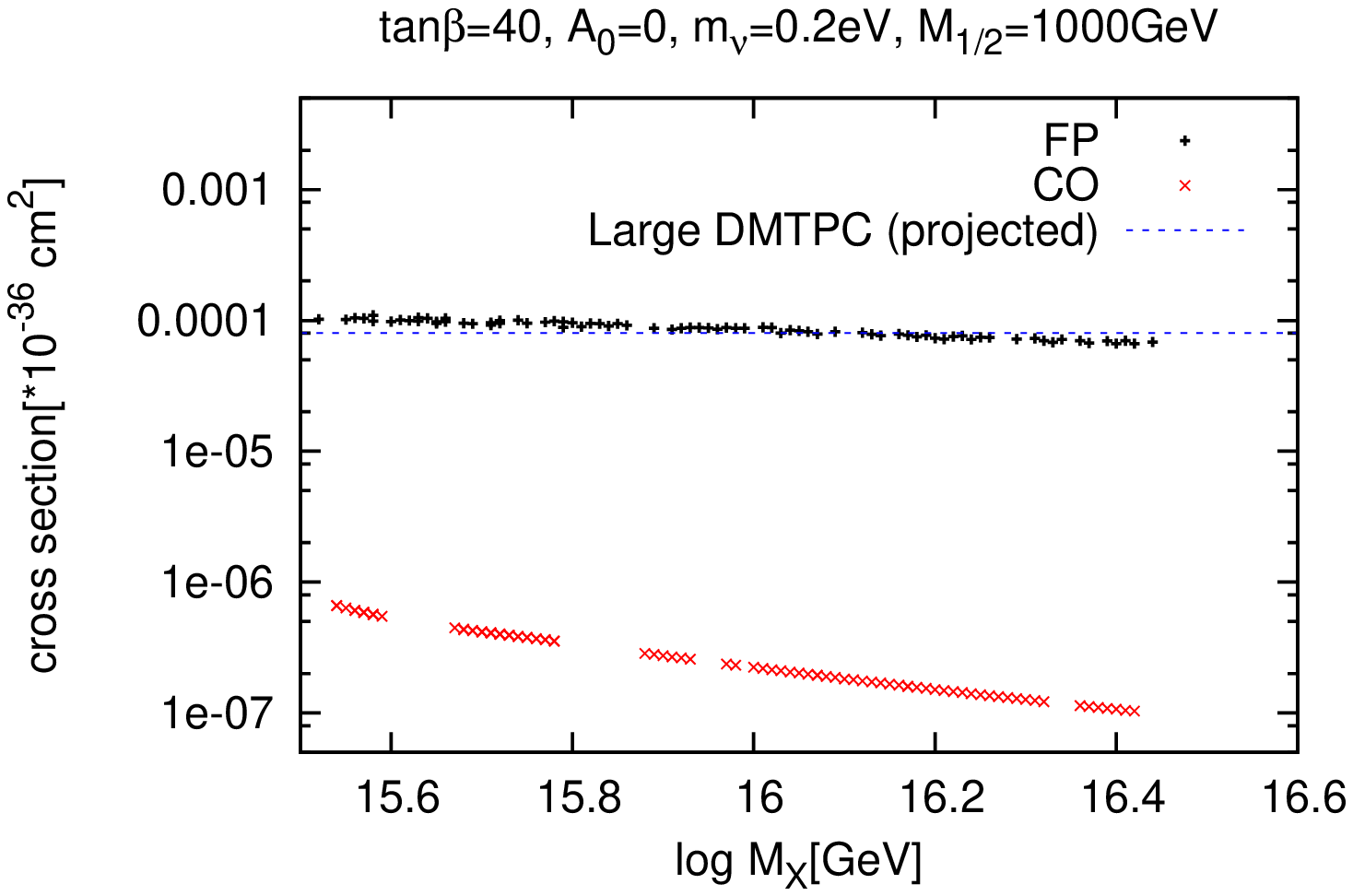}
\caption{Dependence of the SI (left frame) and SD (right frame)
  neutralino--proton scattering cross section on $M_X$, for
  $m_\nu=0.2$ eV. We chose two points that represent the ``focus
  point'' (``FP'') region (black) and the ``co--annihilation''
  (``CO'') region (red or grey). The neutralino mass varies between
  220 GeV and 390 GeV at the smallest and largest $M_X$,
  respectively. The experimental constrains are taken for
  $m_{\tilde \chi_1^0}=220$ GeV, and scale essentially like $m_{\tilde
    \chi_1^0}$.}
    \label{fig:sigmapMx_nu2}
\end{center}
\end{figure}

Figs.~\ref{fig:sigmapMx_nu2} shows the same cross sections for
smaller (heaviest) neutrino mass, $m_\nu = 0.2$ eV, as well as larger
gaugino mass, $M_{1/2} = 1$ TeV. Recall that the smaller $m_\nu$
requires a larger Yukawa coupling $Y_N$, which, among other things,
reduces the weak--scale $\tilde \tau$ masses. This allows to satisfy the
relic density constraint (\ref{eq:wmap}) for two distinct choices of
$m_0$. We continue to call the choice with $m_0^2 \gg M_{1/2}^2$, and
resulting sizable higgsino component of the LSP, the ``focus point''
\cite{focus}, even though the $SO(10)$ model does not show
``focusing'' behavior of any Higgs soft breaking mass
\cite{Drees:2008tc}. In the ``co-annihilation'' region the relic
density is largely determined by $\tilde \chi_1^0 - \tilde \tau_1$
co--annihilation \cite{staucoan} in both mSUGRA and the $SO(10)$
model.

For the focus point, we find the SI cross section to be almost
independent of $M_X$. Note that $m_{\tilde \chi_1^0}$ is now well
above $m_t$ in the entire range of $M_X$ shown. Moreover, $m_A$
increases with decreasing $m_\nu$ \cite{Drees:2008tc}; the decrease of
$m_A$ with decreasing $M_X$ is therefore less pronounced than in
Fig.~\ref{fig:sigmapMx}. Finally, reducing the LSP mass increases the
annihilation cross section, which scales like $m_{\tilde
  \chi_1^0}^{-2}$ away from thresholds. In compensation,
gaugino--higgsino mixing has to be reduced. This reduces the LSP
couplings to neutral Higgs bosons, offsetting the effect of the
reduction of $m_A$ as far as the SI cross section is concerned. In the
SD case, we again observe a slight increase of the cross section with
decreasing $M_X$, as in Fig.~\ref{fig:sigmapMx}, away from the $t \bar
t$ threshold. Note also that, in spite of the increased LSP mass, the
``focus point'' scenario remains easily testable by near--future
direct search experiments, at least via the SI cross section.

On the other hand, for the co--annihilation point, both the SI and SD
cross sections increase by one order of magnitude when $M_X$ is
reduced to its smallest allowed value. Here the Dark Matter relic
density is mainly determined by the mass difference between the LSP
and the lightest stau, which does not strongly depend on
$\mu$. Instead, the correct relic density is obtained through the
direct effect of $m_0$ on $m_{\tilde \tau_1}$. Due to the strong
(exponential) dependence of the relic density on the $\tilde \chi_1^0
- \tilde \tau_1$ mass splitting, only relatively minor adjustments of
$m_0$ are required, which do not lead to significant changes of $\mu$.
In contrast, the additional Yukawa couplings in the $SO(10)$ model
reduce $|\mu|$ all over the parameter space. Therefore, in the
co--annihilation region the larger higgsino component of
$\tilde{\chi}_1^0$ gives rise to larger scattering amplitudes, in
particular via the Higgs-- and $Z^0-$exchange diagrams that dominate
the SI and SD cross sections, respectively. The SI cross section is
enhanced in addition by the decreasing $m_A$. As a result, at the
smallest value of $M_X$ this cross section even approaches the
XENON100 sensitivity.

In order to understand the strong dependence of these cross sections
on $M_X$, one has to keep in mind that reducing $M_X$ increases the
effect of the new Yukawa coupling $Y_N$ in two ways. First, reducing
$M_X$ reduces the intermediate scales $M_R$ and $M_C$ even more,
i.e. $\ln (M_X/M_C)$ and $\ln (M_X/M_R)$ increase when $M_X$ is
decreased. This increases the energy range where this coupling is
effective in the RGE. Secondly, the reduction in $M_R$ has to be
compensated by the increase of $Y_N$ in order to keep the very large
Majorana neutrino mass in the see--saw expression (\ref{seesaw}) constant.

The most robust indirect neutralino Dark Matter detection signal
\cite{jungman} is due to the capture of DM particles by the Sun, which
greatly enhances the neutralino density near the center of the
Sun. Eventually capture and annihilation of neutralinos in the Sun
reach equilibrium. The only annihilation products that can escape
the Sun are neutrinos. In particular, muon neutrinos produce muons via
charged current interactions; these muons can be searched for by
``neutrino telescopes''.

\begin{figure}[!tp]
\begin{center}
\includegraphics[width=8.2cm,height=8.2cm,angle=270]{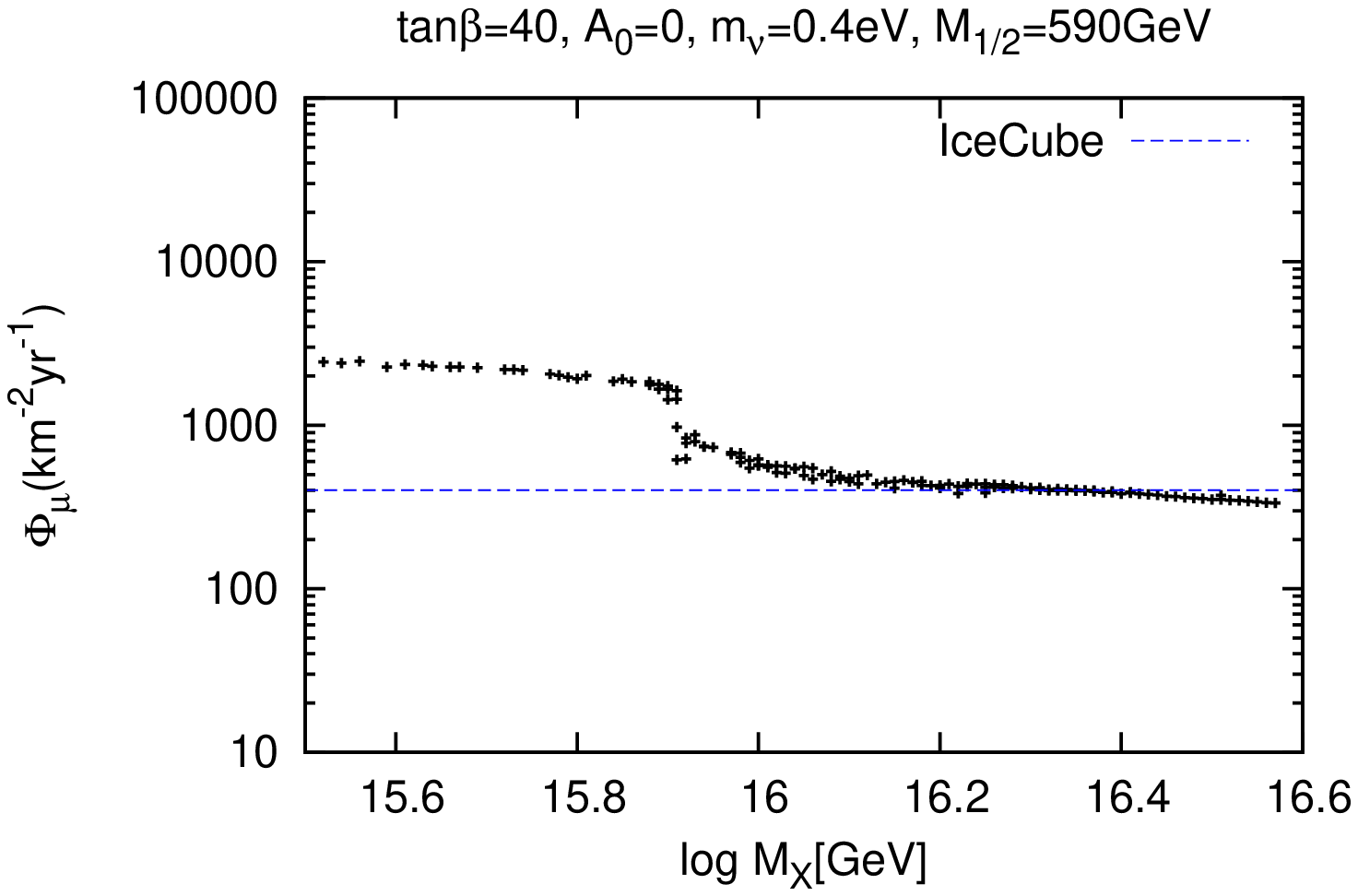}
\includegraphics[width=8.2cm,height=8.2cm,angle=270]{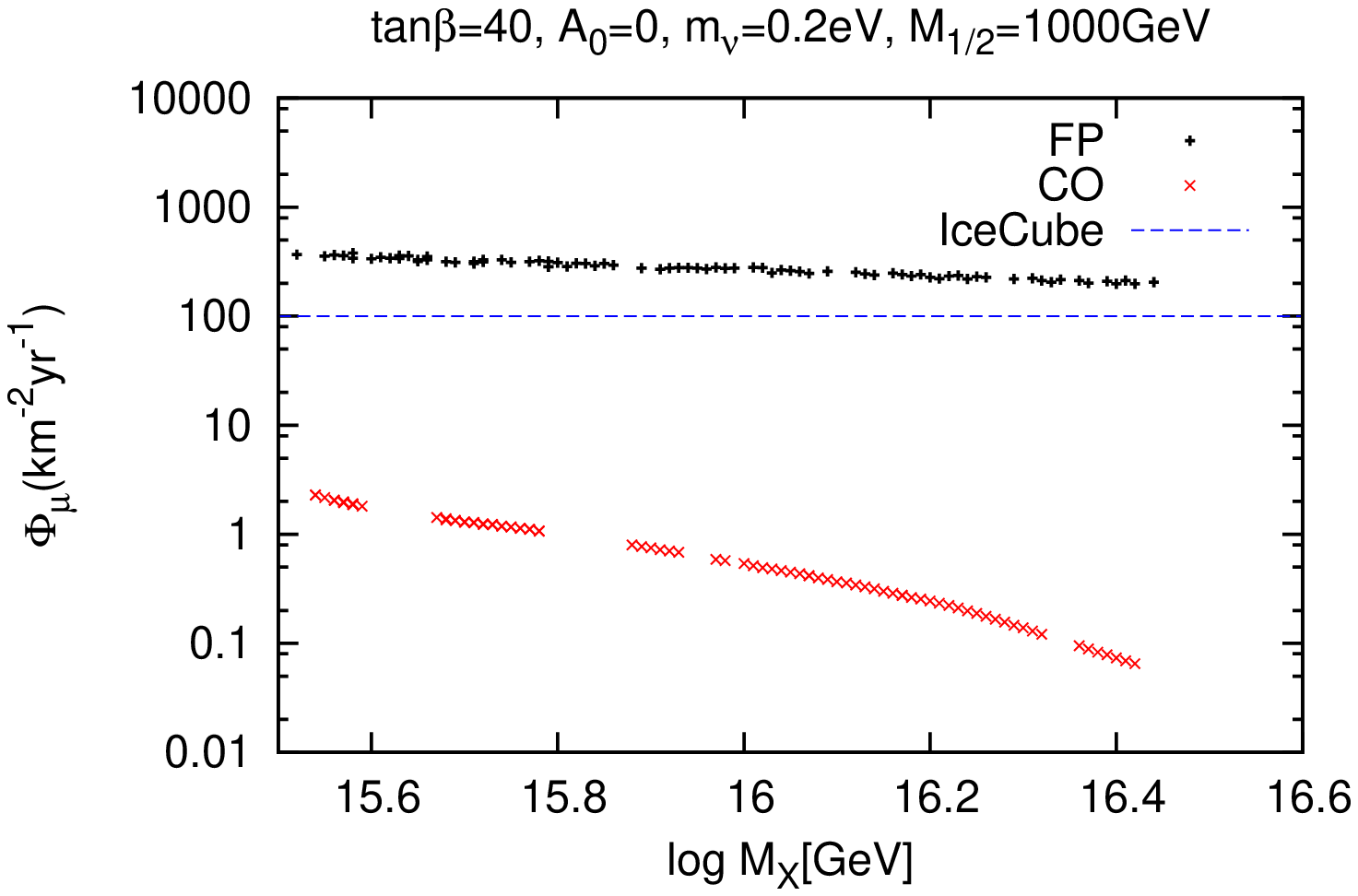}
\caption{Neutrino--induced muon flux from neutralino annihilation in
  the Sun as a function of $M_X$ for $m_\nu=0.4$ eV (left); 
  $m_\nu=0.2$ eV (right). The IceCUBE sensitivity limits are for the smallest
  LSP mass in the respective frames; in the relevant range of masses,
  the sensitivity limit scales roughly $\propto 1/m_{\tilde \chi_1^0}$.}
\label{fig:neutrino}
\end{center}
\end{figure}

In Figs.~\ref{fig:neutrino} we plot the resulting muon flux as function
of $M_X$, and compare it to the ``best case'' sensitivity of IceCUBE
\cite{:2007td}, using the input parameters of Fig.~\ref{fig:sigmapMx}
(left frame) and Fig.~\ref{fig:sigmapMx_nu2} (right). Note that the
overall neutrino flux is essentially fixed by the capture rate. The
neutralinos interact with nuclei in the Sun mostly via Higgs and $Z^0$
exchange. The capture rate is thus again sensitive to the higgsino
components of the mostly bino--like neutralinos. It also depends on
the mass of the neutralinos: the heavier the LSP, the less likely it
is to lose enough energy in the interaction to become gravitationally
bound to the Sun. The predicted muon flux therefore increases faster
with decreasing $M_X$ than the cross sections shown in
Figs.~\ref{fig:sigmapMx} and \ref{fig:sigmapMx_nu2} do.

The muon flux also depends on the (mean) neutrino energy, since the
neutrino charged current cross section increases with
energy. Annihilation into pairs of $W^\pm$ or $Z^0$ bosons leads to
the hardest neutrino spectra, and hence to the largest signals.
Annihilation into $t \bar t$ gives a somewhat softer spectrum, since
some of the energy is taken away by the $b-$quarks.
This enhances the effect of the $t \bar t$ threshold visible in the
left frame: To the left of this threshold, neutralinos predominantly
annihilate directly into massive gauge bosons, while to the right of
the threshold, annihilation into $t \bar t$ dominates.

Of course, the neutrino energy also scales with the mass of the
annihilating neutralinos. Indeed, the sensitivity limit on the muon
flux decreases with increasing LSP mass for $m_{\tilde \chi_1^0} \lsim
500$ GeV \cite{:2007td}. However, in the muon flux itself this effect
is compensated by the reduction of the neutralino flux impinging on
the Sun, which scales like $1/m_{\tilde \chi_1^0}$. Nevertheless, this
effect keeps the expected flux in the ``focus point'' region well
above the sensitivity limit even for the larger value of $M_{1/2}$
chosen in the right frame. However, the flux in the co--annihilation
region remains well below the IceCUBE sensitivity even for the
smallest possible value of $M_X$.

\begin{figure}[ht]
\begin{center}
\includegraphics[width=8.2cm,height=8.2cm,angle=270]{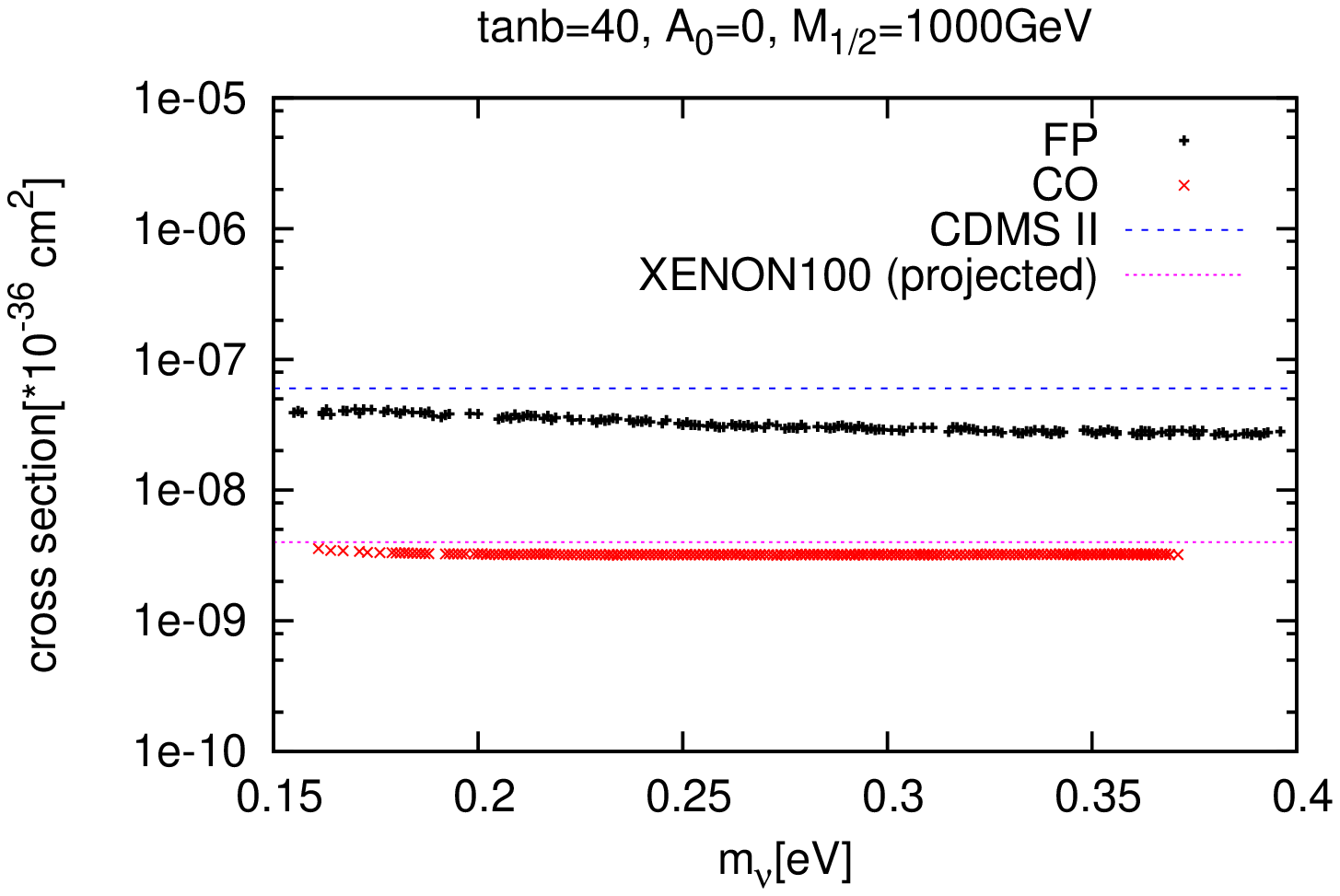}
\includegraphics[width=8.2cm,height=8.2cm,angle=270]{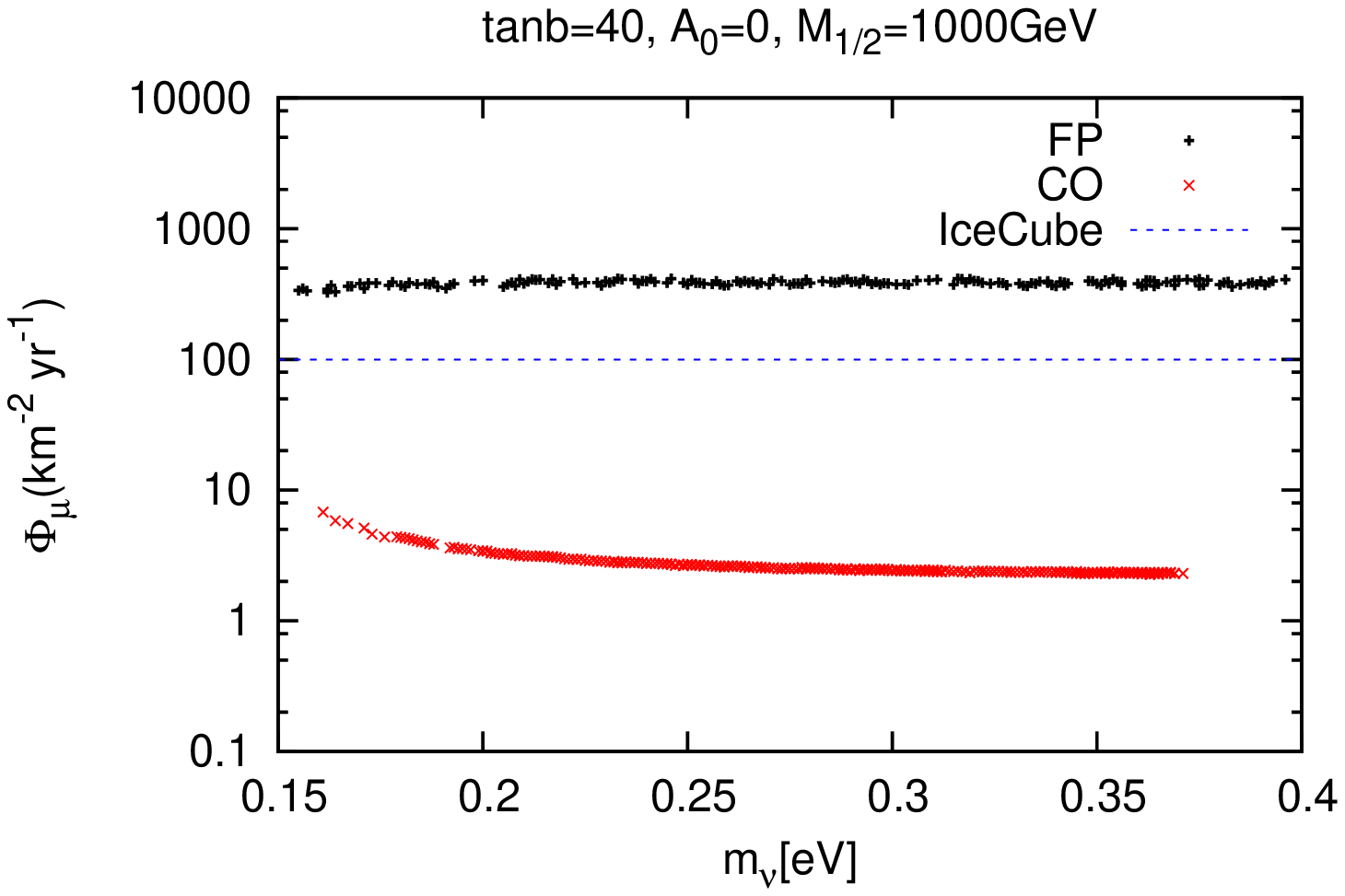}
\caption{SI neutralino--proton scattering cross section (left) and
  neutrino--induced muon flux from neutralino annihilation in the Sun
  (right), as function of $m_\nu$.}
    \label{fig:sigmapMnu}
\end{center}
\end{figure} 

In Fig.~\ref{fig:sigmapMnu}, we show the SI proton--neutralino cross
section as well as the neutrino--induced muon flux as function of
$m_\nu$, for $M_X=3\cdot 10^{15}$ GeV. We see that in the FP region, the
cross section slightly increases with decreasing $m_\nu$. Recall that
decreasing $m_\nu$, i.e. increasing the coupling $Y_N$, reduces
$\mu$. In order to keep the relic density fixed one has to increase
$\mu$ again by decreasing $m_0$, which in turn leads to the decrease of
the Higgs boson masses; this overcompensates the increase of $m_A$
with decreasing $m_\nu$ if all soft breaking parameters are kept
fixed. The reduced Higgs boson masses increase the scattering cross
section. However, it also increases the importance of the $A-$exchange
contribution to the $\tilde \chi_1^0$ annihilation cross section at
rest. For $\tan\beta \gg 1$, $A-$exchange mostly leads to $b \bar b$
final states, which produce very soft neutrinos. This effect
over--compensates the (small) increase in the neutralino capture cross
section, leading to the (very slight) decrease of the muon flux with
decreasing $m_\nu$ in the FP region.

In the co--annihilation region, increasing the Yukawa coupling $Y_N$
reduces $m_{\tilde \tau_R}$ as well as $\mu$. The two effects tend to
cancel, but a net reduction of $m_{\tilde \tau_1}$ results. This has
to be compensated by increasing $m_0$ in order to keep the relic
density in the desired range. This, as well as the effect of $Y_N$ in
the RGE, increases $m_A$. The increase of $m_A$ and the decrease of
$\mu$ essentially cancel in the SI cross section. However, increasing
$m_A$ also reduces the importance of neutralino annihilation to $b
\bar b$. This increases the average neutrino energy, which explains
the slight increase of the muon flux with decreasing $m_\nu$.

We have also computed the near--Earth flux of antiprotons due to the
annihilation of relic neutralinos in the halo of our galaxy. As well
known, the flux depends sensitively on several poorly known
astrophysical quantities. One of these is the density of Dark Matter,
which is reasonably well known ``locally'', but not near the center of
the galaxy, where it is largest. Note that, unlike positrons,
antiprotons can diffuse from the galactic center to Earth. We
illustrate this uncertainty by comparing three different halo
models. The ``N03'' profile has been derived \cite{Edsjo:2004pf}
starting from a profile extrapolated from $N-$body simulations
\cite{Navarro:2003ew}, assuming that baryon infall compresses the Dark
Matter distribution near the galactic center adiabatically. In the
opposite extreme, one can assume that baryon infall heats the dark
halo, leading \cite{Edsjo:2004pf} to a profile similar to the
(phenomenologically apparently quite successful) ``Burkert'' profile
\cite{burkert}. Finally, the ``NFW'' profile \cite{nfw} lies between
these extremes.

\begin{figure}[!bp]
\begin{center}
\includegraphics[width=8.2cm,height=8.2cm,angle=270]{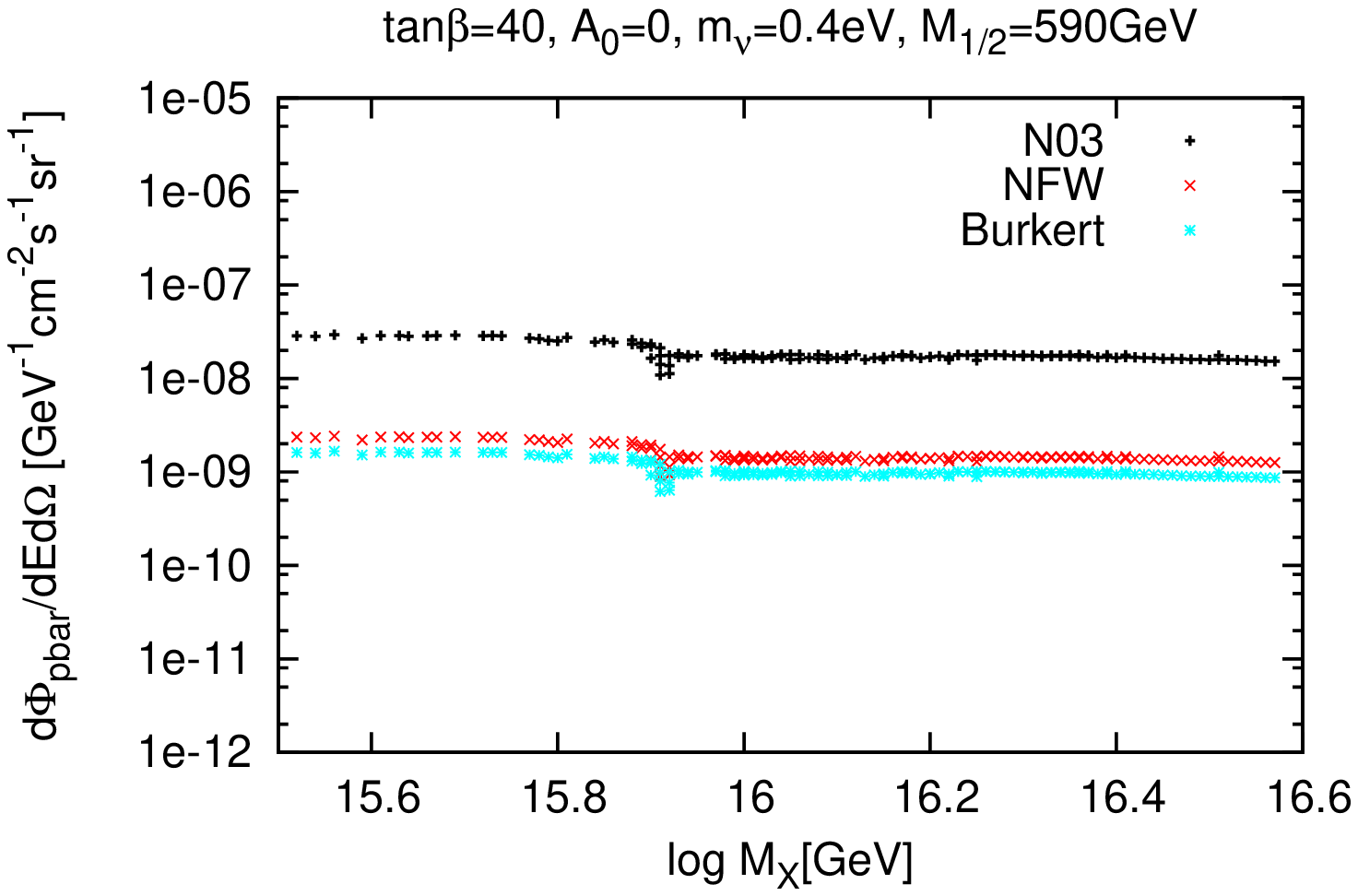}
\includegraphics[width=8.2cm,height=8.2cm,angle=270]{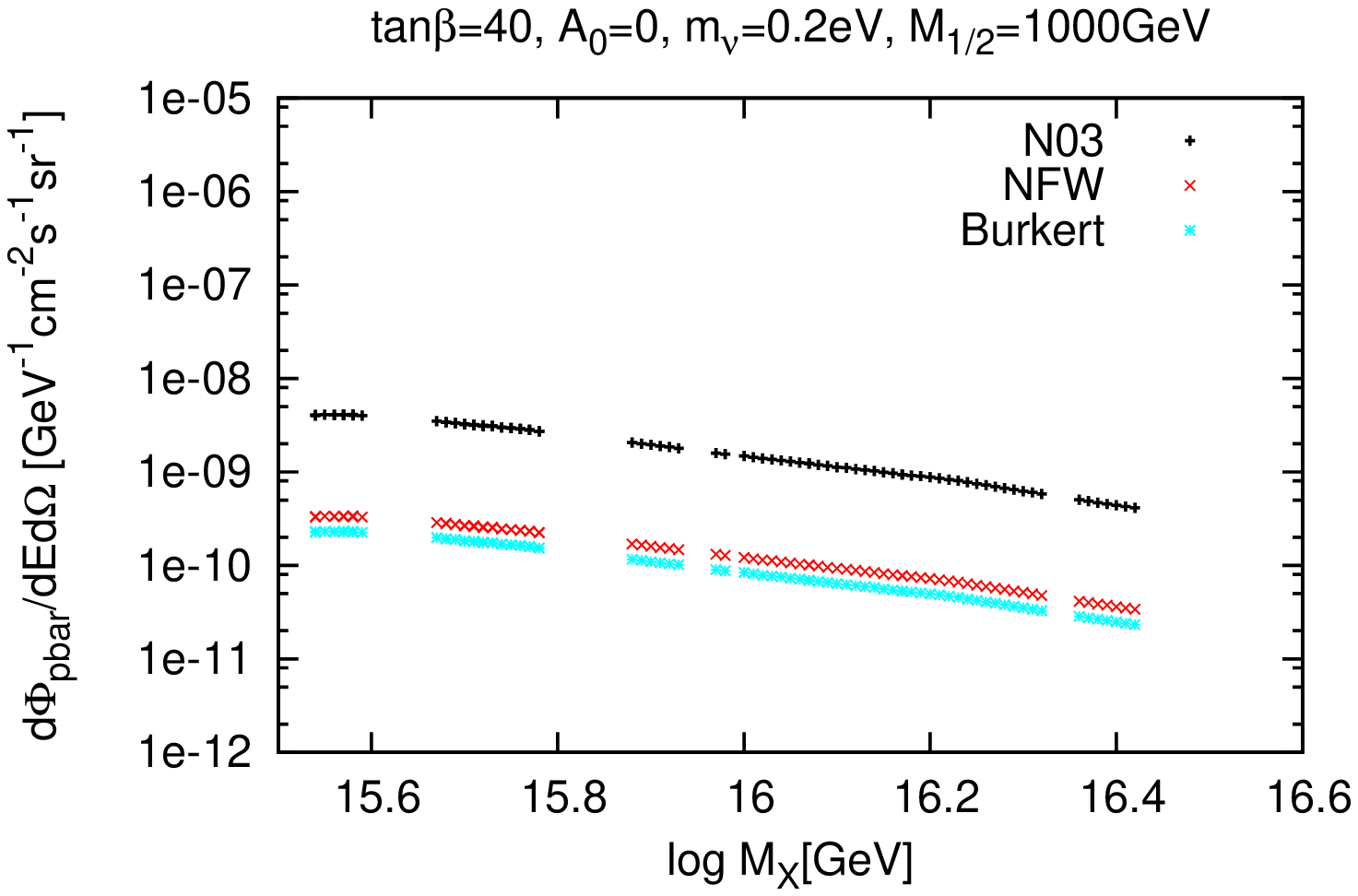}
\caption{Antiproton differential flux for different halo models for
  $m_\nu=0.4$ eV (``focus-point'', left) and $m_\nu=0.2$ eV
  (``co-annihilation'', right).}
    \label{fig:halo}
\end{center}
\end{figure} 

Figs.~\ref{fig:halo} show the dependence of the antiproton flux on
$M_X$. Antiprotons are produced in the Galactic halo due to the
hadronization of antiquarks produced in neutralino Dark Matter
annihilation. As a result, the typical $\bar p$ energy is well below
$m_{\tilde \chi_1^0}$. We show their differential flux at
a kinetic energy 20 GeV, where the signal--to--background ratio is
expected to be optimal \cite{Profumo:2004ty}. We illustrate the
dependence on the halo model using the three profiles discussed above.

The left frame of Fig.~\ref{fig:halo} is for the ``focus point''
region, with small $Y_N$ and relatively small $M_{1/2}$. In this case
the relic density is determined by $\tilde \chi_1^0$ annihilation with
itself, and is dominated by annihilation from the $S-$wave, which is
the only contribution relevant for the $\bar p$ flux. As a result, the
$\tilde \chi_1^0$ annihilation cross section remains essentially
constant in the left frame. However, we saw in
Fig.~\ref{fig:sigmapMx}, where the same parameters were used, that
$m_{\tilde \chi_1^0}$ decreases by nearly a factor of two as $M_X$ is
decreased. This increases the $\tilde \chi_1^0$ annihilation rate,
computed as the product of flux and cross section, by almost a factor
of four. However, decreasing $m_{\tilde \chi_1^0}$ also makes it
increasingly more difficult to produce antiprotons at 20 GeV. As a
result, the $\bar p$ flux near Earth only increases very slightly as
$M_X$ is decreased.

The right frame shows results for a point in the co--annihilation
region, with larger $M_{1/2}$ and smaller $m_\nu$. Here the relic
density is essentially determined by $\tilde \tau_1 - \tilde \chi_1^0$
co--annihilation. The annihilation cross section increases
significantly with decreasing $M_X$, due to the decrease of (almost)
all weak--scale sparticle and Higgs boson masses. Moreover, $m_{\tilde
  \chi_1^0}$ now remains so high that getting 20 GeV antiprotons is
not difficult. As a result, the rate increases by about an order of
magnitude as $M_X$ is reduced to its lower bound.

This seems impressive, but is still smaller than the difference in the
predictions based on the N03 and Burkert profiles. Additional
systematic uncertainties come from the propagation of the antiprotons;
here we have used DarkSUSY default parameters. Note finally that the
$\bar p$ flux that can be inferred from the $\bar p / p$ ratio
measured by the PAMELA satellite \cite{pamela_pbar} and the
well--known \cite{pdg} proton flux is about $2 \cdot 10^{-7}$
GeV$^{-1}$cm$^{-2}$s$^{-1}$sr$^{-1}$, well above even the most
optimistic prediction in Fig.~\ref{fig:halo}. Given that the
prediction for the background also has sizable uncertainties, we
conclude that the observation of cosmic antiprotons is not a very
promising test of the models discussed here.

\begin{figure}[t]
\begin{center}
\includegraphics[width=8.2cm,height=8.2cm,angle=270]{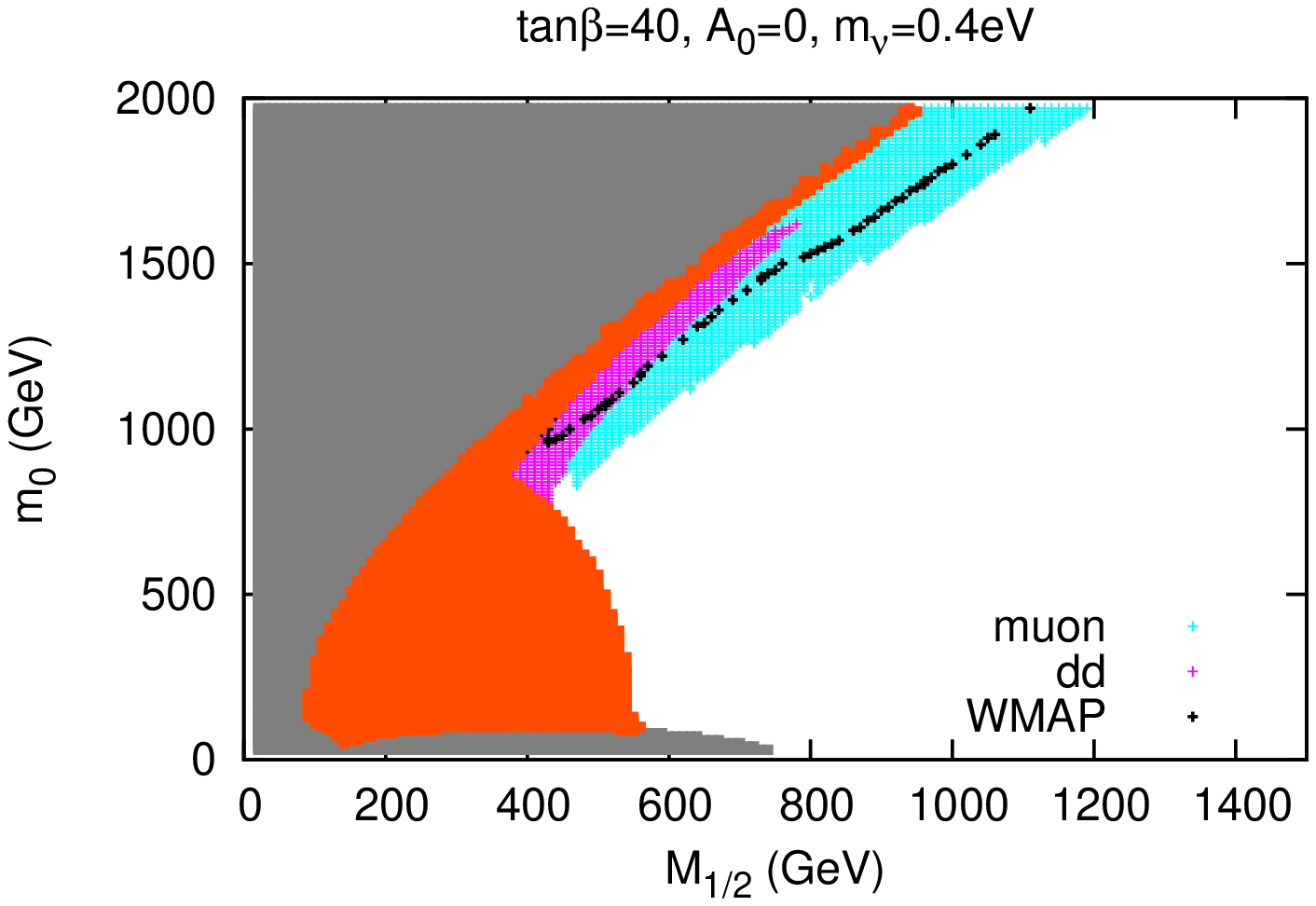}
\includegraphics[width=8.2cm,height=8.2cm,angle=270]{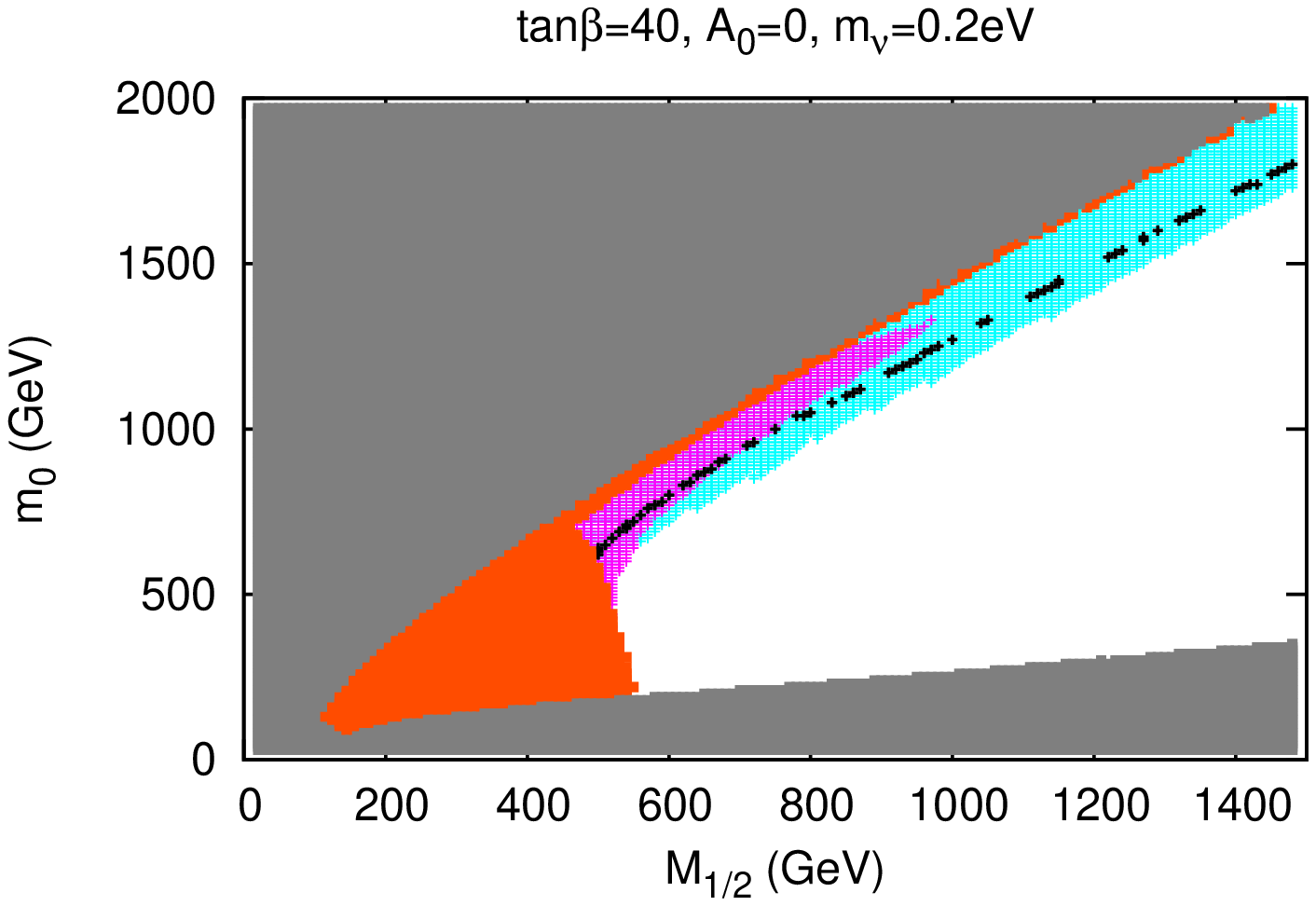}
\caption{Regions in the $(M_{1/2},m_0)$ plane for $m_\nu=0.4$ eV
  (left); $m_\nu=0.2$ eV (right). The grey area is excluded by the
  EWSB condition or by too light sfermions, and the scarlet area is
  excluded by the LEP limits on Higgs and chargino masses. The black
  points give us the correct Dark Matter relic density. The light blue
  region can be tested by searching for muon neutrinos originating
  from $\tilde \chi_1^0$ annihilation in the Sun, while in the magenta
  region, the $\tilde \chi_1^0 p$ scattering cross section exceeds the
  CDMS--II bound. The dependence of the detectability on the WIMP mass
  is taken into account, but we assume a fixed local WIMP density.}
    \label{fig:ps}
\end{center}
\end{figure}

Prospects for direct and indirect DM search in the $SO(10)$ model with
the smallest allowed $M_X$ are summarized in Figs.~\ref{fig:ps}, for
two different values of $m_\nu$. The regions of parameter space that
give the neutrino--induced muon flux from $\tilde \chi_1^0$ annihilation
in the Sun above the IceCUBE sensitivity limit are depicted as light
blue. The regions where the spin--independent neutralino--proton cross
section exceeds the CDMS--II bound are shown in magenta. Note that we
always assume fixed local neutralino density when deriving these
bounds, independent of the value of $\Omega_{\tilde \chi_1^0} h^2$
predicted in standard cosmology. We also show the region excluded by
the electroweak symmetry breaking (EWSB) condition or by too light
sfermions (grey) as well as that excluded by the LEP limits \cite{pdg}
on the masses of Higgs bosons and charginos (scarlet). The black
points are where the Dark Matter relic density satisfies
Eq.(\ref{eq:wmap}).

We find that, as in mSUGRA \cite{Baer:2004qq}, the region of high
$m_0$, where $\tilde \chi_1^0$ has a sizable higgsino component, will
soon be covered by direct searches and also by the neutrino indirect
search. Recall from Figs.~\ref{fig:sigmapMnu} that the region of
parameter space to be probed by XENON100 is much larger than that
probed by IceCUBE. Compared to mSUGRA, for given $M_{1/2}, \ A_0$ and
$\tan\beta$ this region occurs at significantly lower values of $m_0$;
this is true in particular for small $m_\nu$, i.e. sizable Yukawa
coupling $Y_N$ (right frame). Moreover, in this region neutralino Dark
Matter remains detectable out to much larger values of $M_{1/2}$ than
in mSUGRA, since the ratio $m_{\tilde \chi_1^0} / M_{1/2}$ is nearly
two times smaller in our scenario than in mSUGRA.

The co--annihilation region is difficult to see in Figs.~\ref{fig:ps},
since it is very narrow. It extends to $M_{1/2} \simeq 750 \ (1400)$
GeV for $m_\nu = 0.4 \ (0.2)$ eV. Unfortunately this region will not
be tested by near--future Dark Matter search experiments. However, we
saw in Fig.~\ref{fig:sigmapMx_nu2} that the $\tilde \chi_1^0 p$
scattering cross section exceeds that in mSUGRA by about an order of
magnitude. Much of this region will therefore be testable by
ton--scale direct Dark Matter detection experiments.

\section{Collider searches}

In this section, we consider signatures at the LHC for our model. It
remains sufficiently similar to mSUGRA that the overall search
prospects, i.e. the reach in sparticle masses, is essentially the same
in both models; i.e., discovery should be possible out to $m_{\tilde
  g} \sim 3 \ (2)$ TeV for $m_{\tilde q} \simeq \ (\gg) m_{\tilde g}$
\cite{recentbaer}, once the LHC reaches its full energy and
luminosity. Of course, at the smallest allowed value of $M_X$ the
reach in $M_{1/2}$ is nearly two times larger in our model, but
$M_{1/2}$ is not measurable by TeV--scale experiments.  We will
therefore focus on ways to distinguish the model from mSUGRA using
measurements at colliders, with emphasis on the LHC.

\subsection{Benchmark points}

We performed detailed analyses of collider signals for two distinct
benchmark points. The input parameters as well as superparticle and Higgs
spectra are listed in Table~\ref{table:parameter}. We chose points
that satisfy all constraints, including the Dark Matter relic density
constraint (but ignoring the indication of a deviation of the magnetic
dipole moment of the muon from the Standard Model prediction
\cite{gmu}). We chose $M_X$ at its lower bound of $3 \cdot 10^{15}$
GeV, and small $m_\nu = 0.2$ eV, in order to maximize the differences
between our model and mSUGRA. On the other hand, we chose the
parameters of the mSUGRA points such that the sparticle spectra are as
similar as possible to those of the corresponding $SO(10)$ benchmark
points. In particular, we adjust the values of $M_{1/2}$ such that the
gluino masses are essentially the same in both models. Moreover, we
chose the same $m_0$ in both models, since this gives similar first
and second generation squark masses. In this way we hope to isolate
the non--trivial effects of the additional couplings via the RGE.

\begin{table}[t!]
\begin{center}
\begin{tabular}{|l||l|l||l|l|l|}
\hline
parameter & $SO(10)$ 1 & mSUGRA 1 & $SO(10)$ 2 & mSUGRA 2a & mSUGRA 2b\\
\hline
\hline
$M_{1/2}$ & 1100 & 600 & 1000 & 550 & 550 \\
$m_0$ & 1400 & 1400 & 280 & 280 & 280 \\
$A_0$  & 0 & 300 & 0  & -120 & 0\\
$\tan\beta$ & 40 & 52  & 40 & 40 & 41.5\\
\hline 
\hline 
$\mu$  & 307& 587  & 607& 682 & 663 \\
\hline
\hline
$\tilde \chi^0_1$ & 243 & 253 & 229 & 227 & 227\\
$\tilde \chi^0_2$  & 313& 468 & 430 & 431 & 430\\
$\tilde \chi^0_3$ & 317 & 597  & 615& 690 & 671\\
$\tilde \chi^0_4$ & 519 & 618 & 628 & 698 & 680\\
$\tilde \chi^{\pm}_1$ & 298 & 470  & 434& 434 & 433\\
$\tilde \chi^{\pm}_2$  & 517& 615  & 625& 694 & 676\\
\hline
$\tilde{g}$ & 1423 & 1427 & 1246 & 1258 & 1258\\
\hline
$\tilde u_L, \tilde c_L$ & 1865 & 1862 & 1168 & 1178 & 1177\\
$\tilde u_R, \tilde c_R$ & 1842 & 1836 & 1140 & 1140 & 1140\\
$\tilde d_L, \tilde s_L$ & 1870 & 1868  & 1175& 1185 & 1184\\
$\tilde d_R, \tilde s_R$ & 1843 & 1831 & 1138 & 1135 & 1134\\
$\tilde t_1$ & 1205 & 1311 & 874 & 886 & 897\\
$\tilde t_2$ & 1409 & 1495 & 1062 & 1086 & 1088\\
$\tilde b_1$ & 1418 & 1463  & 998& 1016 & 1017\\
$\tilde b_2$ & 1529 & 1532  & 1056& 1074 & 1076\\
\hline
$\tilde e_L, \tilde \mu_L$ & 1490 & 1461 & 544 & 473 & 473\\
$\tilde e_R, \tilde \mu_R$ & 1466 & 1421 & 472 & 354 & 354\\
$\tilde \tau_1$ & 900 & 960 & 238 & 237 & 238\\
$\tilde \tau_2$ & 1230 & 1259 & 488 & 464 & 465\\
\hline
$h^0$ & 116 & 116 & 115 & 115 & 115\\
$H^0, A^0$ & 1018 & 588  & 580& 615 & 593\\
$H^{\pm}$ & 1021 & 594  & 586& 621 & 598\\
\hline
\hline
$\Omega_{\tilde \chi_1^0} h^2$ & 0.09 & 0.09  & 0.11 & 0.11 & 0.12\\
\hline
$P_\tau$ & 0.72 & 0.96 & 0.92 & 0.89 & 0.89 \\
\hline
\hline
\end{tabular}
\end{center}
\caption{Benchmark points used in our analysis of collider
  signals. Mass spectra are  calculated using a modified version of
  \texttt{SOFTSUSY 2.0}. Rows 2 through 5 give the input
  parameters. Row 6 is the weak--scale value of $\mu$. Rows 7 through
  28 give the on--shell mass of the indicated sparticle or Higgs
  boson, while row 29 gives the prediction for the scaled LSP relic
  density. Finally, the last row is the longitudinal polarization of
  the $\tau$ lepton in $\tilde \tau_1^- \rightarrow \tau^- \tilde
  \chi_1^0$ decays. All dimensionful quantities are in GeV.}
\label{table:parameter}
\end{table}

Point 1 is chosen such that, at least in the $SO(10)$ model, the
lightest neutralino has a significant higgsino component. This
requires $m_0 > M_{1/2}$ even in this model. However, for the same
gluino mass, one would need much larger $m_0$ to achieve a similarly
small $\mu$ in mSUGRA. This would put squarks out of the reach of the
LHC, making the scenario easily distinguishable from our $SO(10)$
point. We instead chose to increase $\tan\beta$ from 40 to 52, and
also took a nonvanishing (but fairly small) $A_0$. This leads to
greatly reduced mass of the CP--odd Higgs boson, i.e. we are now close
to the ``$A-$pole'' region \cite{dn3} where $\tilde \chi_1^0$
annihilation is enhanced since $A-$exchange in the $s-$channel becomes
(nearly) resonant. These changes do not affect $\mu$ very much,
i.e. in our mSUGRA point 1 the LSP remains a nearly pure bino.

Benchmark point 2 lies in the co--annihilation region. Recall that the
new coupling $Y_N$ reduces $m_{\tilde \tau_R}$ below its mSUGRA
prediction. Choosing $M_{1/2}$ such that one gets the same $\tilde g$
(or $\tilde \chi_1^0$) mass, while keeping all other input parameters
the same, would thus lead to an mSUGRA point with too high a relic
density. We consider two different methods to correct for this. In
mSUGRA point 2a we take non--vanishing $A_0$, such that $m_{\tilde
  \tau_R}$ is reduced and $\mu$ is increased; the latter also
decreases $m_{\tilde \tau_1}$, helping to get a sufficiently large
$\tilde \chi_1^0 - \tilde \tau_1$ co--annihilation cross section. In
mSUGRA point 2b, this is instead achieved by increasing $\tan\beta$,
which again reduces $m_{\tilde \tau_R}$ and increases $\tilde \tau_L -
\tilde \tau_R$ mixing. Notice that in either case the change of these
input parameters is not very dramatic.

At the $SO(10)$ point 1, all of the squarks as well as the gluino are
significantly heavier than all of the neutralinos and charginos.
Furthermore, due to the low value of $|\mu|$, the heaviest neutralino
$\tilde{\chi}_4^0$ has the largest $SU(2)$ gaugino component. Due to
the small higgsino mass, the dominant SUSY production channel is $q
\bar q \rightarrow \tilde{\chi}_2^0 \tilde{\chi}_1^{\pm}$. Mostly due
to this process, the total inclusive SUSY production cross section at
$\sqrt{s} = 14$ TeV is nearly three times larger than that of the
mSUGRA point 1. However, $\tilde{\chi}_{2,3}^0$ and $\tilde{\chi}_1^+$
decay predominantly into $\tilde \chi^0_1$ and a quark--antiquark
pair, which carries relatively little energy due to the small mass
splitting. Direct $\tilde \chi_{2,3}^0 \tilde \chi_1^\pm$ production
therefore predominantly gives rise to events with four relatively soft
jets, and correspondingly only a small amount of missing $E_T$. This
signal will be completely swamped by backgrounds, e.g. from $W,Z$ plus
multi--jet production. The inclusive cross section for squark and
gluino production, which should be detectable in this scenario (see
below), is quite similar in the $SO(10)$ and mSUGRA versions of point
1.

Another distinctive feature of $SO(10)$ point 1 is the much smaller
polarization $P_\tau$ of $\tau$ leptons produced in $\tilde \tau_1^-
\rightarrow \tau^- \tilde \chi_1^0$ decays. $P_\tau$ depends
\cite{Nojiri:1994it} both on $\tilde \tau_L - \tilde \tau_R$ mixing
and on gaugino--higgsino mixing. In the case at hand, $\tilde \tau_1$
is dominated by the $\tilde \tau_R$ component in both the $SO(10)$
model and in mSUGRA; the $\tilde \tau_L$ component is slightly smaller
in the $SO(10)$ case due to the reduced value of $\mu
\tan\beta$. However, the $SO(10)$ model features much stronger
bino--higgsino mixing in this case. Note that the bino couples $\tilde
\tau_R$ to $\tau_R$, while the (down--type) higgsino couples $\tilde
\tau_R$ to $\tau_L$. As a result, $P_\tau$ is significantly smaller in
the $SO(10)$ case.

$P_\tau$ can be measured via the energies of hadronic $\tau$ decay
products \cite{hm}. Of course, this requires a copious source of
$\tilde \tau_1$ particles. At an $e^+e^-$ collider this measurement
can therefore only be performed if the beam energy is well above
$m_{\tilde \tau_1}$, i.e. $\sqrt{s} \gsim 2$ TeV in our case. Monte
Carlo simulations indicate \cite{Godbole:2004mq} that $P_\tau$ could
then be determined with sufficient accuracy to distinguish these
scenarios. At the LHC this measurement is probably only possible if
$\tilde \tau_1$ particles are produced copiously in the decays of
gluinos and/or squarks \cite{Guchait:2002xh}; this is not the case in
our benchmark point 1.

For the $SO(10)$ point 2, since the gluino and squarks are relatively
lighter, the dominant sparticle production process is $q g \rightarrow
\tilde{g}\tilde{q}_{L,R}$. Our choices of $M_{1/2}$ and $m_0$ ensure
that the corresponding cross section is very similar in the $SO(10)$
and both mSUGRA scenarios. 

Recall that we adjusted the mSUGRA parameters such that we get very
similar $m_{\tilde \tau_1}$, and hence similar LSP relic density, as
that in the $SO(10)$ scenario. These adjustments also imply that the masses
of third generation squarks are only slightly smaller in the $SO(10)$
scenario than in both mSUGRA scenarios, i.e. the effect of the new
Yukawa couplings on sfermion masses has been partly compensated by
adjusting soft breaking parameters. However, the effect of the new
couplings is still visible in $|\mu|$, which is significantly smaller
in the $SO(10)$ benchmark point than in both mSUGRA variants.

Moreover, having adjusted parameters such that we obtain similar
gaugino and first generation squark masses, we get significantly
heavier first generation sleptons in $SO(10)$ than in mSUGRA
\cite{Drees:2008tc}. This can be tested trivially at $e^+e^-$
colliders operating at $\sqrt{s} > 2 m_{\tilde e_R}$. However, even in
the mSUGRA versions of our point 2, sleptons are too heavy for direct
slepton pair production to yield a viable signal at the LHC
\cite{slep_LHC}.

In the next Subsection, we will therefore focus on events containing
charged lepton pairs originating from the decays of squarks and
gluinos. We will show that this allows to distinguish the $SO(10)$
points from their mSUGRA analogues, using the fact that $|\mu|$ is
smaller in the $SO(10)$ scenarios.

\subsection{Measurements using di--lepton events at the LHC}

%\begin{table}[!tp]
%\begin{center}
%\begin{tabular}{|l|l|}
%\hline
%Modes & $SO(10)$ 1 \\
%\hline
%\hline
%$\tilde \chi_2^0 \to \tilde \chi_1^0 e^+e^-$ & 2.7 \% \\
%$\tilde \chi_3^0 \to \tilde \chi_1^0 e^+e^-$ & 3.4 \% \\
%$\tilde \chi_4^0 \to \tilde \chi_3^0 Z^0$ & 21 \% \\
%$\tilde \chi_2^+ \to \tilde \chi_1^+ Z^0$ & 27 \% \\
%$\tilde \chi_2^+ \to \tilde \chi_2^0 W^+$ & 25 \% \\
%$\tilde \chi_2^+ \to \tilde \chi_3^0 W^+$ & 24 \% \\
%$\tilde \chi_1^+ \to \tilde \chi_1^0 e^+\nu_e$ & 11 \% \\
%\hline
%\end{tabular}
%\end{center}
%\caption{Branching ratios for the important modes in $SO(10)$ 1 are
%  calculated with \texttt{ISAJET 7.78} \cite{Paige:2003mg}.}
%\label{table:br1}
%\end{table}

In order to analyze the gaugino--higgsino sector of the theory, we
have to rely on neutralinos and charginos produced in the decays of
squarks and gluinos. Direct production of charginos and neutralinos is
only detectable in purely leptonic final states \cite{trilepton}. In
the case at hand the relevant neutralino and chargino states are quite
massive, and have small leptonic branching ratios, leading to very
small signal rates.

We therefore look for events with several energetic jets in addition
to two or more leptons. To that end, we simulate proton--proton
collisions at the LHC ($\sqrt{s} = 14$ TeV) using \texttt{PYTHIA 6.4}
\cite{Sjostrand:2006za} and the toy detector PYCELL. The detector is
assumed to cover pseudorapidity $|\eta|<5$ with a uniform segmentation
$\Delta \eta = \Delta \phi = 0.1$. We ignore energy smearing, which
should not be important for our analyses. We use a cone jet algorithm,
requiring the total transverse energy $E_T$ summed over cells within
$R=0.4$ to exceed 10 GeV; here $R = \sqrt{(\delta \eta)^2 + (\delta
  \phi)^2}$, where $\delta \eta$ and $\delta \phi$ measure the
deviation in pseudorapidity and azimuthal angle from the jet
axis. $t\bar{t}$ and diboson production are assumed to be the main
Standard Model backgrounds in the di--lepton channels we are interested
in.

We require electrons and muons to have $p_T > 10$ GeV and to be
isolated, i.e. to have less than 10 GeV of additional $E_T$ in a cone
with $R = 0.2$ around them. Also, leptons within $R < 0.4$ of a jet
are not counted. These requirements essentially remove leptons from
the decay of $c$ and $b$ quarks. Finally, we require the invariant
mass of opposite--sign, like--flavor lepton pairs to exceed 20 GeV, in
order to suppress contributions involving virtual photons.

\subsubsection{Point 1}

In this case the difference in $\mu$ between the $SO(10)$ and mSUGRA
scenarios is quite drastic: in the $SO(10)$ case, $\mu$ is only
slightly above $M_1$ and well below $M_2$, leading to $m_{\tilde
  \chi_2^0} \simeq m_{\tilde \chi_3^0} \simeq m_{\tilde \chi_1^\pm}$,
only about 70 GeV above the LSP mass, and well below the masses of the
wino--like $\tilde \chi_2^\pm$ and $\tilde \chi_4^0$ states. In
contrast, in the mSUGRA scenario we have $\mu$ slightly above $M_2$,
leading to wino--like $\tilde \chi_2^0$ and $\tilde \chi_1^\pm$ well
above the LSP.

In order to understand what this means for multi--jet plus di--lepton
signatures, we have to analyze the most important sparticle production
and decay channels. In the case at hand, the most important production
channels (after cuts) are squark pair and associated squark plus
gluino production, where the squarks are in the first generation. Most
squarks will decay into a gluino and a quark here, so that most events
start out as gluino pairs with one or two additional jets.

In the $SO(10)$ version of point 1, nearly all gluinos decay into
$\tilde t_1$ plus top, since this is the only allowed two--body decay
of the gluino. In turn, $\tilde t_1$ decays mostly into $\tilde
\chi_1^+ b$ and $\tilde \chi_{1,2,3}^0 t$. These decays are preferred
by phase space, and because here the {\em lighter} neutralinos and
{\em lighter} chargino are dominantly higgsino--like, and hence couple
more strongly to (s)top (since the top Yukawa coupling is larger than
the electroweak gauge couplings). Leptons can then originate from
semi--leptonic decays of top quarks, from leptonic decays of $\tilde
\chi_1^\pm$ states, and from leptonic decays of $\tilde
\chi_{2,3}^0$. Note that the latter decays, which have branching
ratios near 3\%, can only produce di--lepton pairs with invariant mass
below 70 GeV.

In contrast, in the mSUGRA version of point 1, gluinos can only
undergo three--body decays. Nevertheless decays involving third
generation quarks are strongly preferred, since the $\tilde b_1$ and
$\tilde t_1$ exchanged in $\tilde g$ decay can be nearly
on--shell. The dominant decay modes again involve higgsino--like
states, i.e. $\tilde g \rightarrow \tilde \chi_2^+ b \bar t$ or
$\tilde \chi_{3,4}^0 t \bar t$; due to the larger phase space, the
branching ratio for $\tilde g \rightarrow \tilde \chi_1^+ b \bar t$ is
also significant. The higgsino--like states decay into lighter
gaugino--like states plus a real gauge or Higgs boson. Leptons can
then originate from semi--leptonic top decays, and from the decays of
the $W^\pm$ and $Z^0$ decays produced in the decays of the heavier
neutralinos and both charginos. Note that we do not expect any
structure in the di--lepton invariant mass below $M_Z$ in this case.

We apply the following cuts to suppress the Standard Model background
\cite{Aad:2009wy}:

\begin{itemize}
\item At least four jets with $E_T > 150$ GeV each, at least one of
  which satisfies  $E_T > 300$ GeV. 
\item Missing $E_T > 200$ GeV.
\item Transverse sphericity $S_T > 0.2$.
\item Two charged leptons with opposite sign and same flavor (OSSF).  
\end{itemize}

No SM diboson event passed these cuts, and one $t\bar{t}$ event passed,
for a simulated integrated luminosity of $1 \ {\rm fb}^{-1}$. Note
that our cuts are quite generic, not optimized for our scenario. In
our case, the background can be further suppressed by requiring at
least three tagged $b$ quarks in the event (all the final states we
discussed above have at least four $b$ quarks); by requiring the
presence of additional jets (most events will have at least one hard
jet in addition to the gluino pair, which by itself already produces
at least four jets); and/or by optimizing the numerical values of the cuts
employed. This should allow to extract an almost pure SUSY sample,
without significant loss of signal.

\begin{figure}[ht]
\begin{center}
\includegraphics[width=8.2cm,height=8.2cm,angle=270]{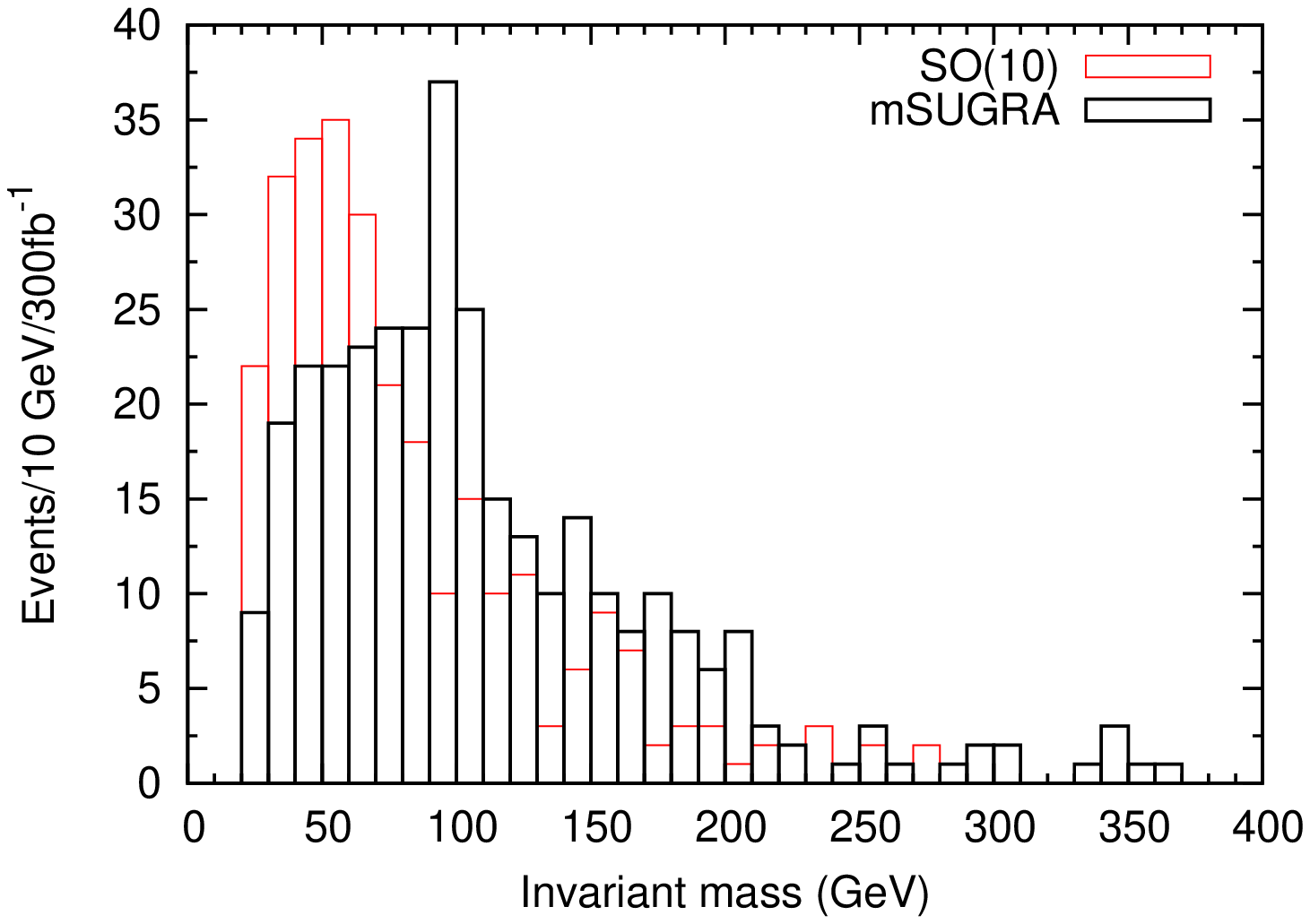}
\includegraphics[width=8.2cm,height=8.2cm,angle=270]{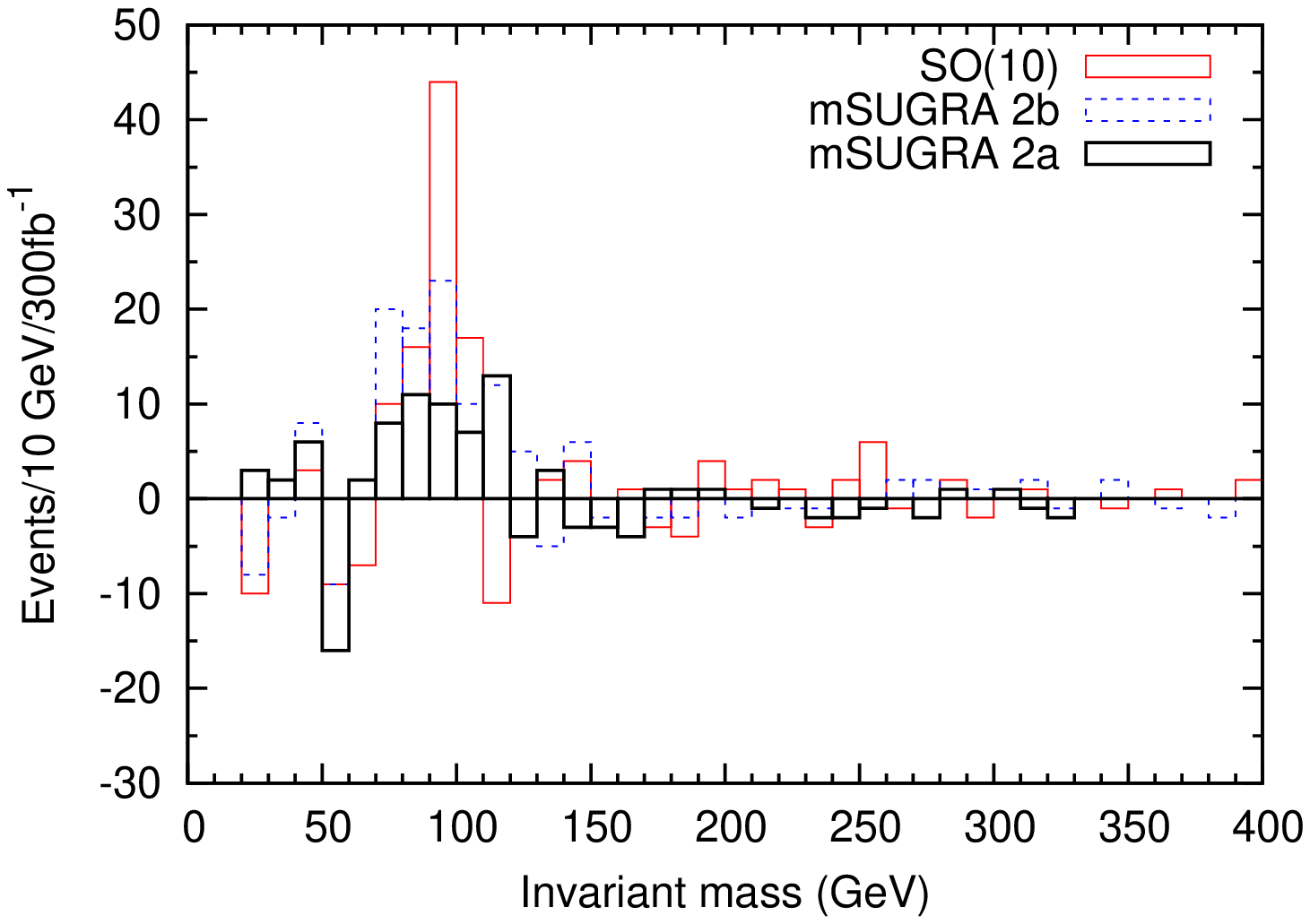}
\caption{Left frame: The dilepton invariant mass distribution of opposite
  sign, same flavor dilepton events after cuts for point 1,
  corresponding to an integrated luminosity of 300 fb$^{-1}$; the black
  and red (grey) histograms are for the mSUGRA and $SO(10)$ versions
  of this point, respectively. Right Frame: The subtracted (OSSF$-$OSOF)
  dilepton invariant mass distribution after cuts for point 2; the
  solid black and dashed blue histograms are for the mSUGRA points 2a
  and 2b, while the solid red (grey) histogram is for the $SO(10)$
  version.}
\label{fig:collider}
\end{center}
\end{figure}

The results of our simulations for point 1 are shown in the left frame
of Figs.~\ref{fig:collider}. We see that the di--lepton invariant mass
distribution peaks near 50 GeV in the $SO(10)$ scenario, whereas it
peaks at $M_Z$ in the mSUGRA scenario. mSUGRA also predicts a somewhat
larger number of events at large di--lepton invariant mass; we saw
above that only in this scenario we expect significant numbers of
on--shell $W^\pm$ bosons from chargino and neutralino decay. The two
distributions should be easily distinguishable, with high statistical
significance, once several hundred fb$^{-1}$ of data will have been
collected.

\subsubsection{Point 2}

We now turn to benchmark point 2. We saw in
Table~\ref{table:parameter} that now $\tilde \chi_2^0 \rightarrow
\tilde e_R^\pm e^\mp$ are allowed in the mSUGRA scenarios, but not in
$SO(10)$. Unfortunately, the branching ratios for these decays remain
at the permille level even in the mSUGRA scenarios. This is partly
due to the small phase space available for these decays, but mostly
due to the fact that $\tilde \chi_2^0$ is dominantly a neutral wino in
the co--annihilation region, and thus has only small couplings to
$SU(2)$ singlet sleptons; recall that $\tilde \tau_1$ also has a
significant $SU(2)$ doublet, $\tilde \tau_L$ component. As a result,
we do not see any evidence for $\tilde \chi_2^0 \rightarrow \tilde
\ell_R^\pm \ell^\mp \rightarrow \tilde \chi_1^0 \ell^\pm \ell^\mp \ (\ell
= e, \ \mu)$ in the mSUGRA scenarios; in particular, no kinematic edge
at $m_{\ell^+\ell^-} = 189$ GeV, the nominal endpoint for this decay
chain, is visible.

We instead first try to find the evidence for increased gaugino--higgsino
mixing in the $SO(10)$ scenario, due to the lower value of $\mu$, by
analyzing the decays of the heavier charginos and neutralinos. In the
case at hand the dominant production channel is associate production
of a first generation squark with a gluino. The decays of first and
second generation squarks will predominantly produce $\tilde
\chi_{1,2}^0$ and $\tilde \chi_1^\pm$ states. On the other hand, here
in the co--annihilation region gluinos are heavier than all squarks,
but -- as usual in scenarios where squark masses unify at some high
scale \cite{dn1} -- gluino decays into third generation quarks and
squarks are preferred. We saw in the discussion of benchmark point 1
that third generation squarks in turn frequently decay into
higgsino--like charginos and neutralinos. We look for these heavier
states through their decays into real $Z^0$ bosons.

\begin{table}[!tp]
\begin{center}
\begin{tabular}{|l|l|l|l|}
\hline
Modes & $SO(10)$ 2 & mSUGRA 2a & mSUGRA 2b \\
\hline
\hline
$\tilde g \to \tilde b_1\bar{b}$ & 12.3 \% & 11.8 \% & 12.0 \% \\
$\tilde g \to \tilde b_2\bar{b}$ & 7.0 \% & 6.2 \% & 6.2 \% \\
$\tilde g \to \tilde t_1\bar{t}$ & 12.3 \% & 12.5 \% & 11.8 \% \\ 
$\tilde g \to \tilde t_2\bar{t}$ & 2.9 \% & - & - \\
$\tilde t_1 \to \tilde \chi_3^0t$ & 18.1 \% & 5.2 \% & 13.6 \% \\
$\tilde t_1 \to \tilde \chi_2^+b$ & 21.9 \% & 21.3 \% & 22.5 \% \\
$\tilde t_2 \to \tilde \chi_3^0t$ & 10.4 \% & 8.0 \% & 8.9 \% \\ 
$\tilde t_2 \to \tilde \chi_2^+b$ & 24.8 \% & 20.4 \% & 22.6 \% \\
$\tilde t_2 \to Z^0 \tilde t_1$ & 7.4 \% & 9.3 \% & 7.5 \% \\
$\tilde b_1 \to \tilde \chi_3^0b$ & 14.6 \% & 10.7 \% & 12.2 \% \\
$\tilde b_1 \to \tilde \chi_2^-t$ & 14.5 \% & 8.7 \% & 9.8 \% \\
$\tilde b_2 \to \tilde \chi_3^0b$ & 14.6 \% & 11.9 \% & 12.9 \% \\
$\tilde b_2 \to \tilde \chi_2^-t$ & 46.1 \% & 39.0 \% & 40.9 \% \\
$\tilde \chi_3^0 \to \tilde \chi_{1,2}^0 Z^0$ & 29.3 \% & 28.0 \% & 28.0 \% \\
$\tilde \chi_2^+ \to \tilde \chi_1^+ Z^0$ & 23.8 \% & 22.8 \% & 22.4 \% \\
\hline
\hline
$\tilde{g} \to Z^0X$ & 7.6\% & 4.3\% & 5.0 \% \\
\hline
\end{tabular}
\end{center}
\caption{Branching ratios for important modes in benchmark point
  2, as  calculated with \texttt{ISAJET 7.78} \cite{Paige:2003mg};
  note that charge conjugate gluino decay modes have to be added. The
  last line denotes the sum of the branching ratios of all gluino
  decay chains which give us a $Z$ boson in the final state.} 
\label{table:br2}
\end{table}

The relevant branching ratios are summarized in
Table~\ref{table:br2}. We see that the slightly reduced third
generation squark masses of the $SO(10)$ scenario increase the
branching ratio for gluino decays into the third generation to 69\% in
the $SO(10)$ case, compared to 61\% (60\%) in mSUGRA point 2a (2b).
Moreover, the decays of third generation squarks into $\tilde
\chi_3^0$ and $\tilde \chi_2^\pm$ are enhanced in the $SO(10)$
case. This is also predominantly a phase space effect; the slightly
reduced squark masses in the $SO(10)$ case are over--compensated by
the reduced masses of the higgsino--like states. This effect is
especially drastic for $\tilde t_1 \rightarrow \tilde \chi_3^0 t$,
where the available phase space volume is very small in the mSUGRA
scenarios. However, the strong phase space dependence of the relevant
partial widths\footnote{If $m_q$ is negligible, the partial width for
  $\tilde q \rightarrow \tilde \chi + q$ is $\propto (1 - m^2_{\tilde
    \chi} / m^2_{\tilde q})^2.$} \cite{msugra} leads to quite
significant differences also in many other modes.\footnote{This also
  holds for decays into $\tilde \chi_4^0$. However, Br$(\tilde
  \chi_4^0 \rightarrow \tilde \chi_{1,2}^0 Z^0)$ only amounts to $\sim
  3\%$. This large difference between the decay modes of the
  higgsino--like states $\tilde \chi_3^0$ and $\tilde \chi_4^0$ can be
  traced back to the fact that $\tilde \chi_3^0$ is a very pure {\em
    symmetric} higgsino, i.e. the higgsino components of this
  eigenvector are nearly equal in both magnitude and sign, whereas
  $\tilde \chi_4^0$ is mostly an {\em antisymmetric} higgsino. As a
  result, the $\tilde \chi_4^0 \tilde \chi_{1,2}^0 Z^0$ couplings are
  suppressed, and the $\tilde \chi_4^0 \tilde \chi_{1,2}^0 h$
  couplings are enhanced, where $h$ is the light neutral Higgs boson.}
Finally, the smaller value of $\mu$ in the $SO(10)$ scenario also
leads to more higgsino--gaugino mixing, and hence to slightly larger
branching ratios for decays of the heavier neutralinos and charginos
into real $Z^0$ bosons. The combination of these three effects leads
to a substantially larger $Z^0$ production rate in gluino cascade
decays in the $SO(10)$ scenario than in mSUGRA.

In order to suppress backgrounds, we first make use of the fact that
most signal events will have (at least) one very energetic jet from
the decay of a first generation squark into a light gaugino--like
state. Moreover, the above discussion shows that many events with real
$Z^0$ bosons in the final state also will have a top quark in the
final state, and/or a real $W^\pm$ boson from $\tilde \chi_1^\pm$
decay; the branching ratio for this latter decay amounts to 12.6\% in
the $SO(10)$ scenario. The decays of these particles can lead to
additional, somewhat softer, jets and/or additional leptons. On the
other hand, since the dominant production channel only contains a
single gluino, we only expect two $b$ (anti--)quarks in the final
state; $b-$tagging will therefore not be of much help to suppress the
$t \bar t$ background, which we again expect to be the most dangerous
one.

This leads us to use two complementary sets of cuts; at the end we
simply add both event samples in order to increase the statistics:
\begin{enumerate}
\item[1)] Set 1
  \begin{itemize}
    \item $E_T(j_1) > 600$ GeV, $E_T(j_2) > 200$ GeV, $E_T(j_3) > 100$
      GeV. 
    \item $N_{\ell} = 2$.
  \end{itemize}

\item[2)] Set 2
  \begin{itemize}
    \item $E_T(j_1) > 300$ GeV, $E_T(j_2) > 150$ GeV, $E_T(j_3) > 75$ GeV.
    \item $N_{\ell} \geq 3$.
  \end{itemize}
\end{enumerate}

Note that we do not apply any cut on missing $E_T$, since this wasn't
necessary to suppress the backgrounds we studied. In Set 1, a modest
missing $E_T$ cut will be needed to suppress the $Z^0 +$jets
background, but this can be done without any significant loss of
signal.

For set 1, we require both charged leptons to be of opposite
charge. In set 2, we chose the opposite--sign lepton pair whose
invariant mass is closer to $M_Z$ as ``$Z^0$ candidate''. We found
that the background from $t\bar{t}$ is almost entirely removed by
either set of cuts. Furthermore, in order to extract the lepton pairs
from the decays of $Z^0$ bosons, we subtract the events with opposite
sign opposite flavor (OSOF) lepton pairs. This removes SUSY
backgrounds where the two leptons originate from independent
(semi--)leptonic decays; in benchmark point 2, these come primarily
from the decays of $W^\pm$ bosons.

The resulting di--lepton invariant mass spectrum is shown in the right
frame of Fig.~\ref{fig:collider}, for an integrated luminosity of 300
fb$^{-1}$. We see that the $Z^0$ peak is indeed much more pronounced
in the $SO(10)$ scenario than in the two mSUGRA scenarios. This allows
to distinguish between $SO(10)$ and mSUGRA at about $3\sigma$
statistical significance in this case.

As noted above, in the $SO(10)$ point 2, we find a branching ratio for
$\tilde \chi_1^\pm \rightarrow W^\pm \tilde \chi_1^0$ of about
12.6\%. In the mSUGRA points 2a and 2b, this branching ratio is only
6.1\% and 6.6\%, respectively. At an $e^+e^-$ linear collider with
sufficient energy to produce $\tilde \chi_1^+ \tilde \chi_1^-$ pairs
this large difference in branching ratios should be straightforward to
measure.

At the LHC we have to pursue a somewhat different strategy: leptonic
decays of these $W^\pm$ can give rise to events with two leptons
of the {\em same} charge (like--sign di--lepton events). These can
originate from associate $\tilde q_L \tilde g$ production where
$\tilde g$ decay also produces a lepton; the charge of this lepton
from gluino decay is uncorrelated to that from squark decay, i.e. half
the time the two leptons will have the same charge. The results of
Table~\ref{table:br2} indicate that the inclusive branching ratio for
$\tilde g \rightarrow \ell^\pm$ is somewhat higher in the $SO(10)$
scenario than in both mSUGRA analogues. Other sources are $\tilde u_L
\tilde u_L$ and $\tilde d_L \tilde d_L$ production, which give rise to
$\ell^+ \ell^+$ and $\ell^- \ell^-$ pairs, respectively, if both
squarks decay into $\tilde \chi_1^\pm$. Of course, gluino pairs can
also produce like--sign dileptons, but the gluino pair production
cross section is relatively small at this benchmark point. Note also
that the physics background for like--sign dileptons events is very
small.

We applied the following cuts to isolate a clean sample of SUSY
events:
\begin{itemize}
\item At least two jets, with $E_T(j_1) > 500$ GeV, $E_T(j_2) > 200$
  GeV. These cuts are quite asymmetric, since we expect at least one
  very energetic jet from $\tilde q_L$ decay in the event.
\item Exactly two equally charged isolated leptons, with $p_T(\ell) >
  10$ GeV as before.
\end{itemize}
With these cuts, we find 492 events in 300 fb$^{-1}$ for the $SO(10)$
scenario, as compared to 422 and 434 events in mSUGRA 2a and 2b,
respectively. This difference of $\sim 3$ statistical standard
deviations is much less than the above discussion would lead one to
expect. The reason is that there is another large source of $\ell^\pm$
from $\tilde q_L$ decay: $\tilde q_L \rightarrow q \tilde \chi_2^0$
with $\tilde \chi_2^0 \rightarrow \tau^\pm \tilde \tau_1^\mp$, and
$\tau^\pm \rightarrow \ell^\pm \nu \bar\nu$.\footnote{The decay of the
  $\tilde \tau_1$ produced in this chain, or via the dominant decay
  $\tilde \chi_1^\pm \rightarrow \tilde \tau_1^\pm \nu$, only produces
  very soft $\tau$ leptons, and hence even softer $\ell^\pm$, since we
  are in the co--annihilation region where the $\tilde \tau_1^\pm -
  \tilde \chi_1^0$ mass difference is small.} Unfortunately the
branching ratios for these decays are somewhat {\em larger} in mSUGRA
than in the $SO(10)$ scenario, because higgsino--gaugino mixing tends
to suppress the corresponding partial widths. This significantly
reduces the difference between the predictions for the total
like--sign dilepton event rate.

One can imagine two strategies to enhance the difference between the
predictions. One possibility is to veto leptons from $\tilde \chi_2^0
\rightarrow \tau^\pm \tilde \tau_1^\mp$ decays by vetoing against the
secondary $\tau^\mp$ from $\tilde \tau_1^\mp$ decay. However, this
$\tau$ will be quite soft, so it is not clear how efficient such a
$\tau$ veto would be. Another possibility is to subtract this source
of hard leptons, by using events with an identified $\tau$ jet and the
known $\tau$ decay branching ratios. Again, the feasibility of this
method depends on $\tau$ tagging efficiencies and their
uncertainties. We do therefore not pursue this strategy any further.

\subsection{Measurements involving Higgs bosons at the LHC}

Our benchmark points have quite large values of $\tan\beta$. This
increases the cross sections for inclusive $gg \rightarrow A,\, H$
production, and for associate $gg \rightarrow b \bar b (A, \, H)$
production. The heavy Higgs bosons can be searched for using their
decays into $\tau^+\tau^-$. According to simulations by the CMS
collaboration \cite{cms}, for $\tan\beta=40$ this would allow
discovery of the heavy Higgs bosons out to $m_A \simeq 650$ GeV with
60 fb$^{-1}$ of data. In particular, we expect a robust signal for
$H,\, A$ production in the mSUGRA version of benchmark point 1, but
not in the $SO(10)$ version. In benchmark point 2, we expect signals
of comparable size in all three cases. The $\tau^+ \tau^-$ invariant
mass resolution should suffice to distinguish between the $SO(10)$ and
mSUGRA 2a scenarios, but distinguishing the $SO(10)$ scenario from
mSUGRA 2b might be challenging.

In benchmark point 2, $\tilde \chi_2^0 \to \tilde \chi_1^0+h$ decays
might also allow to discriminate between $SO(10)$ and the two mSUGRA
analogues. The branching ratio for this decay is 11.5\% in the
$SO(10)$ case, but only 5.3\% (5.7\%) for mSUGRA 2a (2b). About 90\%
of the light Higgs bosons will decay into $b \bar b$ pairs.  Recall,
however, that in this scenario gluino decays frequently lead to $b
\bar b$ pairs in the final state, giving rise to a large SUSY
background. We have therefore not pursued this avenue further.

\section{Summary and Conclusions}

We have investigated the detectability of neutralino Dark Matter and
some LHC signatures for a SUSY$-SO(10)$ model with two intermediate
scales, and compared results with the frequently analyzed mSUGRA
model.

In Sec.~2 we have shown that in the cosmologically allowed region
with large scalar mass parameter $m_0$, the direct detection of Dark
Matter should be possible for the next generation of detectors, at
least for gluino masses up to 2 TeV. In this region of parameter
space, which corresponds to the ``focus point'' region of mSUGRA,
IceCUBE will be able to detect the neutrino-induced muon flux from
neutralino annihilation in the Sun. However, in this region of
parameter space the Dark Matter detection rates predicted in the
$SO(10)$ scenario are quite similar to those in mSUGRA, if one adjusts
parameters such that the physical LSP mass and relic density is the
same in both cases. On the other hand, in the co--annihilation region
prospects for direct Dark Matter detection are much better in the
$SO(10)$ case, although the cross section still remains somewhat below
the projected sensitivity of next generation experiments.
Unfortunately, in this case the neutrino signal from the Sun is
several orders of magnitude below the IceCUBE sensitivity.

We also analyzed the flux of antiprotons from neutralino annihilation
in the halo of our galaxy. Even in the most optimistic case -- with a
large annihilation cross section, and a halo profile strongly peaked
at the galactic center -- the predicted flux falls well below the flux
extracted from the PAMELA $\bar p / p$ flux ratio and the known proton
flux. Given the large uncertainties in both signal and background,
searches for antimatter do not appear to be very promising Dark Matter
search channels in this scenario.

In Sec.~3 we compared some collider signals from the model with those
of mSUGRA. We chose two benchmark points, one of which resembles a
``focus point'' scenario, while the other lies in the co--annihilation
region. We compared those with mSUGRA points with identical gluino and
first generation squark masses, and (almost) identical neutralino
relic density. We showed that events containing hard jets and two (or
more) charged leptons (electrons and muons) can be used to
discriminate between mSUGRA and our $SO(10)$ benchmark points. The
existence and size of a $Z^0$ peak in the di--lepton invariant mass
distribution proved particularly useful. Moreover, in the first
benchmark point the search for heavy neutral MSSM Higgs bosons can
help to discriminate between the $SO(10)$ and mSUGRA models, whereas
in the second point the number of like--sign dilepton events can be
used.

It may be interesting to note that searching for events containing
leptonically decaying $Z^0$ bosons \cite{bbkt} or like--sign dileptons
\cite{barnett} were the first strategies suggested to look for
cascade decays of heavy squarks and gluinos. It was realized later
that other channels with fewer leptons offer better SUSY discovery
reach \cite{oldbaer,recentbaer}. Here we find that $Z^0$ bosons and
like--sign lepton pairs can be very useful for discriminating between
different SUSY models.

In the co--annihilation region, the physical spectra predicted by
$SO(10)$ and mSUGRA are quite similar. The models differ mostly in the
values of $\mu$ derived via electroweak symmetry breaking; the
$SO(10)$ scenario also predicts somewhat reduced masses for third
generation squarks. The fact that our analyses nevertheless led to
significant differences in several signals bodes well for the power of
LHC experiments to distinguish between competing SUSY models. However,
we have not attempted to distinguish the $SO(10)$ model from other
modifications of mSUGRA, e.g. scenarios with non--universal soft
masses for Higgs bosons \cite{nuhm}, which also lead to variations in
$\mu$ as well as the masses of third generation sfermions and Higgs
bosons.

Indeed, it is clearly impossible to distinguish our model from
sufficiently complicated SUGRA scenarios with non--universal scalar
soft breaking masses using superparticle and Higgs boson properties
only, simply because the two models can have identical weak--scale
spectra. Recall, however, that our model also makes other predictions.
For the chosen (extreme) value of the scale of Grand Unification,
proton decay should be within reach of next generation experiments
\cite{pdec}. This low unification scale also implies a rather large
mass for at least one neutrino; this can be probed using cosmological
data \cite{cosmo_nu}, and perhaps even through laboratory experiments
\cite{Drexlin:2005zt}. By combining all the information, we will
hopefully be able to pin down the physics at the Grand Unified scale.

\subsection*{Acknowledgments}
We thank Siba P. Das for helpful discussions on the use of PYTHIA, and
Stefano Profumo for discussions on the detectability of neutralino
Dark Matter by antimatter search experiments. MD and JMK are partially
supported by the Marie Curie Training Research Networks
``UniverseNet'' under contract no. MRTN-CT-2006-035863, and ``UniLHC''
under contract no. PITN-GA-2009-237920. EKP acknowledges the
hospitality of LAPTH and LPSC while part of this work was done.


\begin{thebibliography}{99}

\bibitem{oldso10}
H. Fritzsch and P. Minkowski, Ann. Phys. {\bf 93} (1975) 193;
M.S. Chanowitz, J. Ellis and M.K. Gaillard, Nucl. Phys. {\bf B128} (1977)
506. 

\bibitem{seesaw}
P.~Minkowski, Phys.~Lett.~B {\bf 67} (1977) 421

\bibitem{Drees:2008tc}
M.~Drees and J.M.~Kim,
%``Neutralino Dark Matter in an $SO(10)$ Model with Two-step Intermediate Scale
%Symmetry Breaking,''
JHEP {\bf 0812} (2008) 095, arXiv:0810.1875 [hep-ph].

\bibitem{Aulakh:2000sn}
C.S.~Aulakh, B.~Bajc, A.~Melfo, A.~Rasin and G.~Senjanovic,
%``$SO(10)$ theory of R-parity and neutrino mass,''
Nucl.\ Phys.\ {\bf B597} (2001) 89, hep--ph/0004031.

\bibitem{pdg}
Particle Data Group, C. Amsler et al., Phys. Lett. {\bf B667} (2008) 1.

\bibitem{msugra}
For introductions to supersymmetry in general, and to mSUGRA in particular, see
e.g. M. Drees, R.M. Godbole and P. Roy, {\it Theory and Phenomenology of
  Sparticles}, World Scientific (2004);
H. Baer and X. Tata, {\it Weak scale supersymmetry: From superfields to
  scattering events},  Cambridge, UK University Press (2006).

\bibitem{Allanach:2001kg}
B.C.~Allanach,
%``SOFTSUSY: A C++ program for calculating supersymmetric spectra,''
Comput.\ Phys.\ Commun.\  {\bf 143} (2002) 305, hep--ph/0104145.

\bibitem{micromegas}
G. B\'elanger, F. Boudjema, A. Pukhov and A. Semenov, Comput. Phys. 
Commun. {\bf 176} (2007) 367, hep--ph/0607059.

\bibitem{Gondolo:2004sc}
P.~Gondolo, J.~Edsjo, P.~Ullio, L.~Bergstr\"om, M.~Schelke and E.A.~Baltz,
%``DarkSUSY: Computing supersymmetric dark matter properties numerically,''
JCAP {\bf 0407} (2004) 008, arXiv:astro-ph/0406204;
P. Gondolo, J. Edsj\"o, P. Ullio, L. Bergstr\"om, M. Schelke, E.A. Baltz,
T. Bringmann and G. Duda, http://www.physto.se/~edsjo/darksusy 

\bibitem{jungman}
For a review of neutralino Dark Matter detection, see 
G. Jungman, M. Kamionkowski and K. Griest, Phys. Rep. {\bf 267}
(1996) 195, hep--ph/9506380.

\bibitem{Komatsu2009}
WMAP Collab., E. Komatsu et al., Astrophys. J. Suppl. {\bf 180} (2009)
330, arXiv:0803.0547 [astro--ph].

\bibitem{dmtools}
R. Gaitskell and J. Filippini, http://dmtools.berkeley.edu/limitplots/

\bibitem{Ahmed:2009zw}
CDMS--II Collab., Z.~Ahmed {\it et al.},
%``Results from the Final Exposure of the CDMS II Experiment,''
arXiv:0912.3592 [astro--ph.CO].

\bibitem{focus}
J.L. Feng, K.T. Matchev and T. Moroi, Phys. Rev. {\bf D61} (2000) 075005,
hep--ph/9909334;
J.L. Feng, K.T. Matchev and F. Wilczek, Phys. Lett. {\bf B482} (2000) 388,
hep--ph/0004043.

\bibitem{staucoan}
J.R. Ellis, T. Falk, and K.A. Olive, Phys. Lett. {\bf B444} (1998) 367,
hep--ph/9810360;
J.R. Ellis, T. Falk, K.A. Olive and M. Srednicki, Astropart. Phys. 
{\bf 13} (2000) 181, Erratum-ibid. {\bf 15} (2001) 413, hep--ph/9905481;
M.E. Gomez, G. Lazarides and C. Pallis, Phys. Rev. {\bf D61} (2000) 123512,
hep--ph/9907261.

\bibitem{:2007td}
IceCUBE Collab., arXiv:0711.0353 [astro--ph].
%``The IceCUBE Collaboration: contributions to the 30th International Cosmic
%Ray Conference (ICRC 2007),''

\bibitem{Edsjo:2004pf}
J.~Edsjo, M.~Schelke and P.~Ullio,
%``Direct versus indirect detection in mSUGRA with self-consistent halo
%models,''
JCAP {\bf 0409} (2004) 004, astro--ph/0405414.

\bibitem{Navarro:2003ew}
J.F.~Navarro {\it et al.},
%``The Inner Structure of LambdaCDM Halos III: Universality and Asymptotic
%Slopes,''
Mon.\ Not.\ Roy.\ Astron.\ Soc.\  {\bf 349} (2004) 1039, astro--ph/0311231.

\bibitem{burkert}
A. Burkert, Astrophys. J. {\bf 447} (1995) L25.

\bibitem{nfw}
J.F. Navarro, C.S. Frenk and S.D.M. White, Astrophys. J. {\bf 462} (1996)
563, and Astrophys. J. {\bf 490} (1997) 493.

\bibitem{Profumo:2004ty}
S.~Profumo and P.~Ullio,
%``The role of antimatter searches in the hunt for supersymmetric dark
%matter,''
JCAP {\bf 0407} (2004) 006, hep--ph/0406018.

\bibitem{pamela_pbar}
PAMELA Collab., O. Adriani et al., Phys. Rev. Lett. {\bf 102} (2009) 051101,
arXiv:0810.4994 [astro--ph].

\bibitem{Baer:2004qq}
H.~Baer, A.~Belyaev, T.~Krupovnickas and J.~O'Farrill,
%``Indirect, direct and collider detection of neutralino dark matter,''
JCAP {\bf 0408} (2004) 005, hep--ph/0405210.

\bibitem{recentbaer}
For a recent detailed assessment of the mSUGRA discovery potential
of LHC experiments, see H. Baer, V. Barger, A. Lessa and X. Tata,
JHEP {\bf 0909} (2009) 063, arXiv:0907.1922 [hep--ph].

\bibitem{gmu}
See e.g. M. Davier, arXiv:1001.2243 [hep--ph];
T. Teubner, K. Hagiwara, R. Liao, A.D. Martin and D. Nomura,
arXiv:1001.5401 [hep--ph]; and references therein.

\bibitem{dn3}
M. Drees and M.M. Nojiri, Phys. Rev. {\bf D47} (1993) 376,
hep--ph/9207234.

\bibitem{Nojiri:1994it}
M.M.~Nojiri,
%``Polarization of $\tau$ lepton from scalar $\tau$ decay as a probe of
%neutralino mixing,''
Phys.\ Rev.\  D {\bf 51} (1995) 6281, hep--ph/9412374.

\bibitem{hm}
B.K. Bullock, K. Hagiwara and A.D. Martin, Nucl. Phys. {\bf B395}
(1993) 499.

\bibitem{Godbole:2004mq}
R.M.~Godbole, M.~Guchait and D.P.~Roy,
%``Using tau polarization to discriminate between SUSY models and  determine
%SUSY parameters at ILC,''
Phys. Lett. {\bf B618} (2005) 193, hep--ph/0411306.

\bibitem{Guchait:2002xh}
M.~Guchait and D.P.~Roy,
%``Using $\tau$ polarization as a distinctive SUGRA signature at LHC,''
Phys. Lett.  {\bf B541} (2002) 356, hep--ph/0205015.

\bibitem{slep_LHC}
F. del Aguila and L. Ametller, Phys. Lett. {\bf B261} (1991) 326;
H. Baer, C.-h. Chen, F. Paige and X. Tata, Phys. Rev. {\bf D49} (1994)
3283, hep--ph/9311248.

\bibitem{trilepton}
H. Baer, C.-h. Chen, F. Paige and X. Tata, Phys. Rev. {\bf D50} (1994)
4508, hep--ph/9404212.

\bibitem{Sjostrand:2006za}
T.~Sj\"ostrand, S.~Mrenna and P.~Skands,
%``PYTHIA 6.4 Physics and Manual,''
JHEP {\bf 0605} (2006) 026, hep--ph/0603175.

\bibitem{Aad:2009wy}
ATLAS Collab., G.~Aad {\it et al.},
%``Expected Performance of the ATLAS Experiment - Detector, Trigger and
%Physics,''
arXiv:0901.0512 [hep-ex].

\bibitem{dn1}
M. Drees and M.M. Nojiri, Nucl. Phys. {\bf B369} (1992) 54.

\bibitem{Paige:2003mg}
F.E.~Paige, S.D.~Protopopescu, H.~Baer and X.~Tata, hep--ph/0312045.
%``ISAJET 7.69: A Monte Carlo event generator for p p, anti-p p, and e+ e-
%reactions,''

\bibitem{cms}
CMS Collab., A.~De Roeck {\it et al.}, {\it CMS Physics Technical
  Design Report}, Vol. II, Sec.~11.3; see
http://cdsweb.cern.ch/record/942733/files/lhcc-2006-021.pdf

\bibitem{bbkt}
H. Baer, V.D. Barger, D. Karatas and X. Tata, Phys. Rev. {\bf D36}
(1987) 96.

\bibitem{barnett}
R.M. Barnett, J.F. Gunion and H.E. Haber, Phys. Lett. {\bf B315}
(1993) 349, hep--ph/9306204.

\bibitem{oldbaer}
H. Baer, C.-h. Chen, F. Paige and X. Tata, Phys. Rev. {\bf D52} (1995)
2746, hep--ph/9503271, and Phys. Rev. {\bf D53} (1996) 6241,
hep--ph/9512383. 

\bibitem{nuhm}
M. Drees, Y.G. Kim, M.M. Nojiri, D. Toya, K. Hasuko and T. Kobayashi, 
Phys. Rev. {\bf D63} (2001) 035008, hep--ph/0007202;
J.R. Ellis, T. Falk, K.A. Olive and Y. Santoso, Nucl. Phys. {\bf
  B652} (2003) 259, hep--ph/0210205;
H. Baer, A. Mustafayev, S. Profumo, A. Belyaev and X. Tata, JHEP {\bf
  0507} (2005) 065, hep--ph/0504001.

\bibitem{pdec}
See e.g. K. Nakamura, Front. Phys. {\bf 35} (2000) 359; 
A. Bueno {\it et al.}, JHEP {\bf 0704} (2007) 041, hep--ph/0701101.

\bibitem{cosmo_nu}
See e.g. S. Hannestad and Y.Y.Y. Wong, JCAP {\bf 0707} (2007) 004,
astro--ph/0703031.

\bibitem{Drexlin:2005zt}
G.~Drexlin for the KATRIN Collab.,
%``KATRIN: Direct measurement of a sub-eV neutrino mass,''
Nucl.\ Phys.\ Proc.\ Suppl.\  {\bf 145} (2005) 263.

\end{thebibliography}
\end{document}